\documentclass[onecolumn,notitlepage,showpacs,pre,floatfix]{revtex4-1}

\usepackage[utf8]{inputenc}
\usepackage{amsmath}
\usepackage{amssymb}
\usepackage{lipsum}
\usepackage{bm}
\usepackage{graphicx}
\usepackage{graphics}
\usepackage[dvipdf]{epsfig}
\usepackage{subcaption}
\usepackage{float}
\usepackage{wrapfig}
\usepackage{url}
\usepackage{epstopdf} 
\usepackage{xcolor}

\newcommand{\ddt}[2]{\frac{\mathrm{d^2}#1}{\mathrm{ d}#2^2}}
\newcommand{\ddx}[2]{\ddt{}x}
\newcommand{\ddy}[2]{\ddt{}y}
\newcommand{\ddz}[2]{\ddt{}z}

\begin{document}
\title{Role of Gaussian curvature on local equilibrium and dynamics \\ of smectic-isotropic interfaces}

\author{Eduardo Vitral and Perry H. Leo}
\affiliation{Department of Aerospace Engineering and Mechanics,
    University of Minnesota, 110 Union St. SE, Minneapolis, MN 55455, USA}
\author{Jorge Vi\~nals}
\affiliation{School of Physics and Astronomy,
    University of Minnesota, 116 Church St. SE, Minneapolis, MN 55455, USA}

\begin{abstract}
Recent research on interfacial instabilities of smectic films has shown
unexpected morphologies that are not fully explained by classical local
equilibrium thermodynamics. Annealing focal conic domains can lead to conical
pyramids, changing the sign of the Gaussian curvature, and exposing smectic
layers at the interface. In order to explore the role of the Gaussian curvature
on the stability and evolution of the film-vapor interface, we introduce a phase
field model of a smectic-isotropic system as a first step in the study. Through
asymptotic analysis of the model, we generalize the classical condition of local
equilibrium, the Gibbs-Thomson equation, to include contributions from surface
bending and torsion, and a dependence on the layer orientation at the interface.
A full numerical solution of the phase field model is then used to study the
evolution of focal conic structures in smectic domains in contact with the
isotropic phase via local evaporation and condensation of smectic layers. As in
experiments, numerical solutions show that pyramidal structures emerge near the
center of the focal conic owing to evaporation of adjacent smectic planes and to
their orientation relative to the interface. Near the center of the focal conic
domain, a correct description of the motion of the interface requires the 
additional curvature terms obtained in the asymptotic analysis, thus
clarifying the limitations in modeling motion of hyperbolic surfaces solely 
driven by mean curvature.  
\end{abstract}
\date{\today}


\maketitle


\section{Introduction}

Deviations in local intensive thermodynamic variables at curved surfaces determine the forces that govern their motion outside of thermodynamic equilibrium. Equilibrium at curved surfaces~\cite{re:rowlinson82}, initially studied to address capillary phenomena at fluid interfaces, has subsequently played a key role in broad classes of  moving boundary problems. Notable examples include nucleation theory and curvature driven growth in phase transformations~\cite{re:gunton83}, grain growth~\cite{re:mullins56}, sintering of ceramics~\cite{barsoum2002fundamentals}, crystal growth in metal alloys, semiconductor, and high temperature superconducting materials~\cite{re:godreche92,onuki2002phase}, including dendritic growth~\cite{re:langer80}, polymer~\cite{re:strobl00} and protein crystal growth~\cite{re:delucas89}, or the related field of pattern formation in Geochemical systems~\cite{re:ortoleva93}. More recently, attention has shifted to more complex physico-chemical and biological systems in which interfaces and the phases they bound include complex constituents and interactions, and often spontaneously broken symmetries. The interplay between microscopic processes and mesoscopic shape is much richer and difficult to elucidate.

Interfacial curvature effects, and in particular those related to the Gaussian curvature, are under active investigation in the emerging field of shape engineering of surfaces and interfaces. The goal is to leverage interfacial curvature distributions to affect controllable and reversible changes in surface morphology, or to use curved substrates to control crystalization, defect formation, and motion. Examples include shape control through the application of external stimuli~\cite{mostajeran2015curvature}, the use of curvature to localize defects and control hierarchical bending, buckling or folding of multilayered surfaces~\cite{re:stoop15,jimenez2016curvature}, the control of fracture by constraining elastic sheets to adopt fixed curvature distributions~\cite{mitchell2017fracture}, or nucleation and growth~\cite{re:gomez15} and elastic instabilities~\cite{re:meng14} on curved surfaces.

Interfacial geometry, and hence interfacial energy, are described by the local mean and Gaussian curvatures, $H$ and $G$ respectively. The mean curvature has been the quantity of primary physical interest in expressing interfacial energy, as it is directly related to the change in interfacial area for a small displacement of the interface. The classical manifestation of this result is the Gibbs-Thomson equation, which relates the change in chemical potential $\delta \mu$ relative to planarity to the mean curvature as $\delta \mu = 2H \sigma_h$, where $\sigma_h$ is the thermodynamic excess free energy (surface tension for a fluid interface). Indeed, this equation has been central to all studies of equilibrium morphology and interfacial motion. If the interface is endowed with its own elasticity, the additional energy is described by the Canham-Helfrich free energy functional~\cite{re:kamien02}, with dependence on $H^{2}$ and $G$, and coefficients given by the so called bending moduli. The simpler case in which the Gaussian curvature term is omitted is known as the Willmore problem~\cite{willmore1996riemannian}. 

Phase field models have been introduced as a convenient and versatile mathematical description of complex interfacial morphologies. The use of phase fields or Ginzburg-Landau type equations in the study of interfacial motion originates from the pioneering work of Cahn and Hilliard~\cite{re:cahn58} of a two phase interface described by an assumed gradient free energy functional. This work was later extended by Allen and Cahn~\cite{re:allen79} who showed that the method could be used to study the unstable motion of a two phase interface outside of thermodynamic equilibrium. The classical result involving motion driven by mean interfacial curvature emerges from the Allen-Cahn equation as the singular limit in which the width of the initially diffuse interface is taken to zero~\cite{re:bronsard91}. The methodology has been subsequently generalized to the case of a conserved order parameter~\cite{re:kawasaki82}, to interfaces separating fluid phases~\cite{re:kawasaki83,re:chen00}, and to interfaces bounding phases that are modulated in equilibrium~\cite{re:elder92}. We introduce a phase field description of a two phase interface separating a modulated and a disordered phase, and examine the resulting thermodynamic relations in the macroscopic limit of a thin interface. Our results lead to an extended Gibbs-Thomson relation governing local equilibrium at a distorted interface that depends not only on its mean curvature but also on interfacial bending and torsion, and on the alignment of the modulated phase with respect to the interface.

Although our results apply generally to modulated-isotropic interfaces, the particular geometries that we investigate numerically are motivated by recent experiments on smectic-A (SmA) films~\cite{kim2011smectic,kim2016controlling}. In SmA liquid crystals, rod like molecules are organized in planes with a distinct inter-layer spacing. When thin films of a liquid crystal in its smectic phase are deposited on substrate, antagonistic boundary conditions (the smectic layers align  perpendicularly to the substrate but parallel to the air interface) induce the liquid layers to bend into conical defects, which forms a periodic array of toroidal focal conic domains (FCDs) on the film surface. Sintering (i.e., reshaping of a SmA at elevated temperatures for an amount of time, with subsequent cooling) of FCDs have shown that the curvature driven evaporation-condensation of smectic layers results in a variety of transient film structures, including conical pyramids, concentric rings and domes. The interplay between mean and Gaussian curvatures in the FCD is key to the complex instabilities and film morphologies under heat treatment that are observed in the experiments.

Smectic films displaying arrays of FCDs constitute a potential platform for surface engineering through heat treatment. Indeed, arrays of focal conics are being investigated as building-blocks for soft lithography patterning~\cite{yoon2007internal,kim2010self}, base structures for the fabrication of superhydrophobic films~\cite{kim2009fabrication}, guides for the self-assembly of nanoparticles~\cite{milette2012reversible,pratibha2010colloidal}, and optically selective microlens photomasks~\cite{kim2010optically}, which make for an efficient way to produce patterns through photolithography. However, there is very limited understanding at present of the role of curvatures on the thermal processes and stability of these arrays, which would be fundamental to fine tune the morphology and properties of resulting patterns. 

A complete transport model of an isothermal smectic film of the type described above requires consideration of an appropriate smectic order parameter, as well as mass and momentum conservation relations. We will focus here on the simpler case of an smectic-isotropic interface, which is sufficient to obtain equilibrium conditions at a distorted interface, and the kinetic equation for the interface that follows from the relaxation of smectic fluctuations. Direct isotropic to smectic transitions are predicted in systems with sufficiently large intermolecular anisotropic interactions~\cite{mederos1989molecular}, and have been observed in a number of systems including virus-polymer mixtures, liquid crystalline polymers and elastomers~\cite{dogic2001development,olbrich1995fluctuations}. Similarly to a smectic-air interface, the smectic-isotropic interface involves smectic layers that are parallel to the interface, and we will choose boundary conditions on a substrate so that smectic layers are perpendicular to it. Therefore, our model system also presents stationary toroidal focal conic configurations.

In Sec.~\ref{sec:model} we briefly summarize the phase field model used and its relation to the more common description based on the smectic layer displacement field. Section~\ref{sec:asymp} studies weakly nonlinear solutions of the model, including the one dimensional, stationary smectic-isotropic profile at coexistence, and the amplitude equation for weakly distorted smectic layers. The amplitude equation helps us to derive analytic equations for the interface without dealing with the oscillatory nature of the smectic layering, and through it we construct a front solution connecting smectic and isotropic phases. With this result we derive a generalized Gibbs-Thomson and interface velocity equations, and find that these equations are different depending on whether the smectic planes are parallel to the interface, or perpendicular (as in exposed smectic layers). In Sec.~\ref{sec:results}, we present our numerical results for a three dimensional configuration in order to verify both stationary solutions and our asymptotic results.  We also examine kinetic phenomena that are not restricted to weak interfacial curvatures.  Starting from a toroidal focal domain, we show how curvature induced evaporation and condensation of SmA planes leads to morphological change and the formation of conical pyramids. Away from regions of large curvature or interfacial cusps, surface evolution is well described by the generalized Gibbs-Thomson equation. In some cases mean curvature driven growth is sufficient to describe interface motion, whereas in others, Gaussian and mean curvature terms are both needed to fully describe interfacial motion.


\section{Model}
\label{sec:model}

The smectic phase of a liquid crystal has uniaxial symmetry: a layered structure along one direction, and liquid like properties along the two transverse directions. We describe such a phase~\cite{chaikin2000principles} with a scalar order parameter $\psi(\mathbf{x},t)$, function of the three dimensional space $\mathbf{x}$ and time $t$, that also accounts for an isotropic phase when its value is zero. At a microscopic scale on the order of the smectic layer separation, the two phase interface is not sharp, but rather has a finite characteristic width which is larger than the smectic layer wavelength. The free energy associated with the order parameter is~\cite{re:brand01,sakaguchi1996stable}

\begin{equation}
  {\cal F}_s \;=\; \int d {\bf x} \; \frac{1}{2}
            \bigg\{ \epsilon \psi^2 
            + \alpha\left[ \left(q_0^2 + \nabla^2 \right)\psi \right]^2            
            - \frac{\beta}{2} \psi^{4} + \frac{\gamma}{3} \psi^{6}\;\bigg\} \; .
  \label{eq:pf-energy}
\end{equation}
A similar functional is found in Amundson and Helfand {\cite{amundson1993quasi}} to study lamellar block copolymer microstructures, based on the Hamiltonian derived by Leibler {\cite{leibler1980theory}} for composition patterns in weak segregation using mean-field theory. Such polymers present the same translational and rotational symmetries as a SmA. The free energy in both cases will be affected in an analogous way when the molecular planes are distorted (by splay or elongation). The liquid crystal elastic moduli are proportional to the coefficient $\alpha$: the term associated with $\alpha$ in the previous energy is the one influenced by distortions, as it penalizes the energy when the SmA layers move away from a parallel alignment with constant interlayer spacing, where $q_{0}$ is the layer wavenumber. The advantage of adopting a phase field model for interface problems in modulated phases is the regularization it introduces, which allows for topological changes to occur smoothly and to dynamically deal with macroscopic singularities.

The coefficients $\alpha$, $\beta$ and $\gamma$ are three constant, positive parameters, and $\epsilon$ is a small bifurcation parameter that describes the distance away from the SmA-isotropic transition temperature. The constants $\beta$ and $\gamma$ are chosen to give a triple well energy (smectic layers and isotropic phase). Although the temperature does not explicitly shows in this form of free energy, it can be adjusted through $\beta$ and $\gamma$ in the sense that they change the relative stability of the smectic and isotropic phases. The term proportional to $\psi^{6}$ is necessary for coexistence between isotropic and smectic phases~\cite{sakaguchi1996stable}, which occurs at the coexistence point $\epsilon_{c} = 27 \beta^{2}/160 \gamma$. For $\epsilon > \epsilon_{c}$, the equilibrium phase is isotropic, $\psi = 0$, whereas for $\epsilon < \epsilon_{c}$, the smectic phase $\psi \approx \frac{1}{2} [A\,e^{i{\bf q}\cdot{\bf x}} + c.c.]$ is in equilibrium. Here $\| \mathbf{q} \| \approx q_{0}$, where $\mathbf{q}$ has an arbitrary orientation.

Spatially localized and periodic states are found not only exactly at $\epsilon_c$, but in a neighborhood of $\epsilon_c$ that grows as $\epsilon_c$ increases~\cite{sakaguchi1996stable}. This is due to a frustration effect~\cite{burke2006localized}, as for $\epsilon$ just above $\epsilon_c$ there is compression of the localized states with respect to the wavelength at $\epsilon_c$, while for $\epsilon$ just bellow $\epsilon_c$ there is a stretching of the localized states. Beyond this neighborhood, the front between the two solutions will move towards either the isotropic or smectic phase.

We consider relaxational evolution of the order parameter away from equilibrium to be solely driven by free energy minimization,

\begin{equation}
\partial_t \psi \;=\; -\frac{\delta {\cal F}_s}{\delta \psi} \;=\; -\epsilon \,\psi 
    - \alpha \, (q_0^2 + \nabla^2)^2 \psi + \beta \,\psi^3 - \gamma \,\psi^5.
\label{eq:pf-dynamics}
\end{equation}
The model defined by Eqs.~(\ref{eq:pf-energy}) and~(\ref{eq:pf-dynamics}) forms the basis of our analytic and numerical analyses described below. It is rotationally invariant, and allows tracking of arbitrarily distorted smectic planes, as well as smectic-isotropic fronts.

A more common description of weakly distorted smectic phases is in terms of the layer displacement field away from a reference planar configuration $u(\mathbf{x},t)$. The order parameter and displacement field descriptions coincide when there is a preferred direction of the smectic planes, and for weak distortions away from planarity. This is accomplished by defining smectic layers as the surfaces of constant phase of $\psi$. For reference layers perpendicular to the $z$ direction, a weakly distorted smectic plane is $\psi = \frac{1}{2}(A e^{i q_{0}(z-u({\bf x},t))} + {\rm c.c})$. In this limit, the free energy follows from the Oseen-Frank energy and is given by~\cite{oseen1933theory,frank1958liquid,degennes1995physics,santangelo2005curvature}

\begin{equation}
    {\cal F}_d \;=\; \int d{\bf x} \left\{ \frac{K}{2}(c_1+c_2)^2
    +\bar{K} c_1 c_2+\frac{B}{2}(\partial_z u)^2 \right\},
\label{eq:of-energy}
\end{equation}
where $c_{1}$ and $c_{2}$ are the two principal curvatures of the layer surface characterized by a constant $\phi({\bf x}) = z - u({\bf x}) = (\pi/q_0)m$, with $m$ being an integer that orders the layering. This surface has a normal ${\mathbf n} = (-\partial_x u,-\partial_y u,1)$ to first order in the distortion.  The constant $K$ is the splay modulus of the liquid crystal, $\bar{K}$ is the so called saddle-splay modulus, and $B$ is the compressibility modulus. Note that the splay term is associated with an energy contribution coming from the mean curvature $H = \frac{1}{2}(c_1+c_2)$, while the saddle-splay is connected to the contribution from the Gaussian curvature $G = c_1c_2$ to the energy. 

It is possible to relate parameters in Eq.~(\ref{eq:pf-energy}) to the Oseen-Frank constants of Eq.~(\ref{eq:of-energy})~\cite{amundson1993quasi}. Consider a longitudinal distortion field $u({\bf x}) = \delta z$, with $\delta \ll 1$. From Eq.~(\ref{eq:of-energy}), the resulting Oseen-Frank free energy density is $f_d = \frac{1}{2} B \delta^2$. Then, by computing the change in free energy through Eq.~(\ref{eq:pf-energy}),  where we take $f'_s$ to be the free energy density for a distorted $\psi({\bf x}') = A\, \mathrm{cos} [q_0 (z+\delta z)]$ and subtracting the undistorted free energy $f_s$, one finds $\Delta f_s = \delta^2 \alpha q_0^4 A^2$. Therefore $B = 2\alpha q_0^4 A^2$. Similarly, by considering a transverse distortion field $u({\bf x}) = \delta \mathrm{cos}(Q x)$ and $u({\bf x}) = \delta [\mathrm{cos}(Q_x x) + \mathrm{cos}(Q_y y)]$, one can compute the change in free energy density according to Oseen-Frank and the phase field model. In the limit of small distortions, one finds that $K = \frac{1}{2}\alpha q_0^2 A^2$ and $\bar{K} = 0 $. Even though it would be required to consider higher order distortions to find an expression connecting $\bar{K}$ to the phase field model parameters, we note that the saddle-splay term in Eq.~(\ref{eq:of-energy}) is a null Lagrangian, and from the Gauss-Bonnet theorem it follows that the energy contribution of this term depends only on the topology of the smectic domain and boundary conditions~\cite{didonna2002smectic}.


\section{Local equilibrium thermodynamics and kinetics of weakly perturbed smectic layers}
\label{sec:asymp}

Before presenting a fully numerical study of the evolution of toroidal focal domains in Sec.~\ref{sec:results}, we discuss in this section the equilibrium conditions at a weakly curved smectic-isotropic front (the Gibbs-Thomson equation), and the equation of motion for the front. Both can be derived from an asymptotic expansion of Eqs.~(\ref{eq:pf-energy}) and~(\ref{eq:pf-dynamics}) about the isotropic to smectic transition point. Our analysis serves to both generalize the classical Gibbs-Thomson equation, and to verify the numerical calculations of Sec.~\ref{sec:results} for fronts that have small curvature and are away from singularities. We also seek to understand how the orientation of the smectic layers with respect to the interface affects these equilibrium conditions, and how this is related to the experimentally observed nonequilibrium structures~\cite{kim2016controlling}. We first use a multiple scale expansion to derive an amplitude equation for Eq.~(\ref{eq:pf-dynamics}) near two phase coexistence, such that we can describe the interface between the two phases without the oscillatory behavior of the order parameter associated with the modulated phase. We then obtain a particular solution of the amplitude equation that corresponds to a planar and stationary front connecting bulk regions of smectic and isotropic phases. Third, we extend this calculation to curved fronts by projecting the amplitude equation into a local frame on the curved front, and derive both the chemical potential and law of motion as a function of front curvatures alone.


\subsection{Weakly nonlinear analysis}

A weakly nonlinear expansion valid near the smectic-isotropic transition is
introduced to describe the slow relaxation of modulated configurations. We set
$\epsilon$ to be a small expansion parameter, and conduct a standard multiple
scale analysis~\cite{manneville1995dissipative,cross1993pattern}. Here $\epsilon
> 0$ since our study lies in the region where both $\psi=0$ and periodic $\psi$
solutions are linearly stable. The order parameter $\psi$ is expanded in powers
of $\epsilon$ as $\psi(\mathbf{x},t)  = \epsilon^{1/4} \psi_{1} + \epsilon^{3/4}
\psi_{2} + \epsilon^{5/4} \psi_{3} \ldots$, and slow spatial and temporal
variables are introduced according to $X=\epsilon^{1/4}x$, $Y=\epsilon^{1/4}y$,
$Z=\epsilon^{1/2} z$ and $T= \epsilon \, t$. The weakly nonlinear analysis will
capture smectic-isotropic fronts when the amplitude of the order parameter in
the smectic phase is small.

Since we are interested in analytic results for the front when the phases are close to coexistence, and $\epsilon_{c} = 27 \beta^{2}/160 \gamma$, one needs to account for the scaling of the parameters $\beta$ and $\gamma$. We follow Sakaguchi and Brand~\cite{sakaguchi1996stable}, so that we fix $\gamma = 1$ and let $\beta$ control the width of the coexistence region. Therefore, $\beta$ must scale as $\beta \sim {\cal O}(\epsilon^{1/2})$, and numerically we will only vary $\epsilon$ and $\beta$ in order to control the structure of the triple well energy. The resulting expansion of Eq.~(\ref{eq:pf-dynamics}) is solved order by order in $\epsilon$. Note that the powers of $\epsilon$ in the expansion of $\psi$ come from the fact that the amplitude of the oscillatory phase is given by $\sqrt{3\beta / 5 \gamma}$, and that we collect terms coming from the expansion of a Laplacian and a biharmonic operator. At ${\cal O}(\epsilon^{1/4})$ we obtain the equation defining the stationary and one dimensional solution in the bulk smectic phase, $\psi_{1} = \frac{1}{2} \left[ A e^{i q_{0} z} + {\rm c.c.} \right]$. At order $\epsilon^{5/4}$ a solvability condition appears that leads to an equation for the amplitude $A$, which when written in the original $\mathbf{x}$ and $t$ variables, reads (details are given in Appendix~\ref{sec:amp-eqn}),

\begin{equation}
  \partial_t A \;=\; -\epsilon A + 4\alpha q_0^2 \partial_z^2 A
                   - 4\,i\,\alpha q_0 \partial_z\nabla^2_{xy} A
                - \alpha \nabla^4_{xy} A
                  + \frac{3}{4}\beta |A|^2A - \frac{5}{8}\gamma |A|^4A
  \label{eq:amp-gen}
\end{equation}
where $\nabla^2_{xy} = \partial_x^2+\partial_y^2$ and $\nabla^4_{xy} = \nabla^2_{xy} \cdot \nabla^2_{xy}$. This amplitude equation is accurate up to terms of ${\cal O}(\epsilon^{5/4})$. Even though this equation was derived for small $\epsilon$, we will later show numerically that it remains accurate for finite values of this parameter. In our simulations we use $\epsilon_c \gtrsim 0.5$ in order to have a coexistence region of finite width, sufficient for stable numerical computation~\cite{sakaguchi1996stable}, and also to have a sufficiently large range of $\epsilon$ to perform thermal treatment studies.

The amplitude equation can be written in variational form as $\partial_{t} A = -
\delta {\cal F}_{A}/\delta A^{*}$, where $A^{*}$ is the complex conjugate of
$A$, and the associated free energy is,

\begin{equation}
  {\cal F}_{A}\, [A,A^*] \;=\; \int d{\bf x}
  \;\bigg[\; \alpha|(2 q_0\partial_z - i \nabla^2_{xy})A|^2
  +\epsilon |A|^2-\frac{3}{8}\beta|A|^4 
    + \frac{5}{24}\gamma|A|^6 \;\bigg].
  \label{eq:amp-energy}
\end{equation}
Equation~(\ref{eq:amp-energy}) describes up to ${\cal O}(\epsilon^{5/4})$ the relaxation of slowly varying bulk smectic modulations. The relationship between the parameters of the phase field and Oseen-Frank free energies can be obtained from the energy ${\cal F}_A$ as well. In terms of a small displacement $u$, we can write $A = \frac{1}{2}|A|e^{-iq_0 u}$ and similarly for the complex conjugate $A^*$. By substituting into Eq.~(\ref{eq:amp-energy}),  we obtain the compressibility term as $\alpha q_0^4 |A|^2 |\partial_z u|^2$, which when compared to the Oseen-Frank free energy leads to $B = 2\alpha q_0^4 |A|^2$. Also from this substitution we obtain $\frac{1}{4} \alpha q_0^2 |A|^2 |\partial_x^2 u + \partial_y^2 u|^2$ for the splay part, and hence $K = \frac{1}{2}\alpha q_0^2 |A|^2$.


\subsection{Stationary, one dimensional, smectic-isotropic front}

The amplitude equation, Eq.~(\ref{eq:amp-gen}), describes the relaxation of
weakly distorted smectic planes. Near coexistence, however, it can also be used
to describe a continuous front solution connecting smectic and isotropic
regions. In order to find such a one dimensional solution $A = A(z)$ for a
planar front perpendicular to the $z$ direction, we substitute $A = |A|
e^{i\phi}$ into Eq.~(\ref{eq:amp-gen}), where $\phi$ is the phase of the complex
amplitude. The stationary phase equation leads to
\begin{equation*}
    \partial_z (|A|^2\partial_z\phi) = 0, \quad {\rm so}\:{\rm that} \quad |A|^2\partial_z\phi = {\rm constant}.
\end{equation*}
Since $|A| = 0$ for the isotropic phase (at $z \rightarrow \infty$) and $|A|$
has a constant value in the smectic phase ($z \rightarrow - \infty$), this
implies that $\partial_z\phi = 0$. The equation for the amplitude
$|A|$ ($A$ for simplicity) becomes independent of the phase and is given by,
\begin{equation}
    -\epsilon A + 4\alpha q_0^2\partial_z^2 A
    + \frac{3}{4}\beta A^3 - \frac{5}{8}\gamma A^5 \;=\; 0
    \label{eq:amp-par}
\end{equation}
The constant amplitude $A$ in the smectic phase is 
\begin{equation}
    A^2 \;=\; \frac{3\beta + \sqrt{9\beta^2-40\epsilon\gamma}}{5\gamma}
\label{eq:amp-sma}
\end{equation}
By denoting $A = A_p(z)$, Eq.~(\ref{eq:amp-par}) can be solved to yield a planar smectic-isotropic front exactly at $\epsilon = \epsilon_c$, given by,

\begin{equation}
	A_p(z) = \sqrt{\frac{18\beta}{5\gamma}} \,\bigg[ 4 + \textrm{exp} \bigg( \pm 
	\frac{z - z_0}{2 \sqrt{\alpha_s/3}} \bigg) \bigg]^{-1/2} .
    \label{eq:amp-sol}
\end{equation}
The front is centered around $z_0$ (arbitrary) and has width proportional to $\alpha_s = 40\alpha\gamma/9\beta^2$. 

\begin{figure}[h]
\centering
\includegraphics[width=0.5\textwidth]{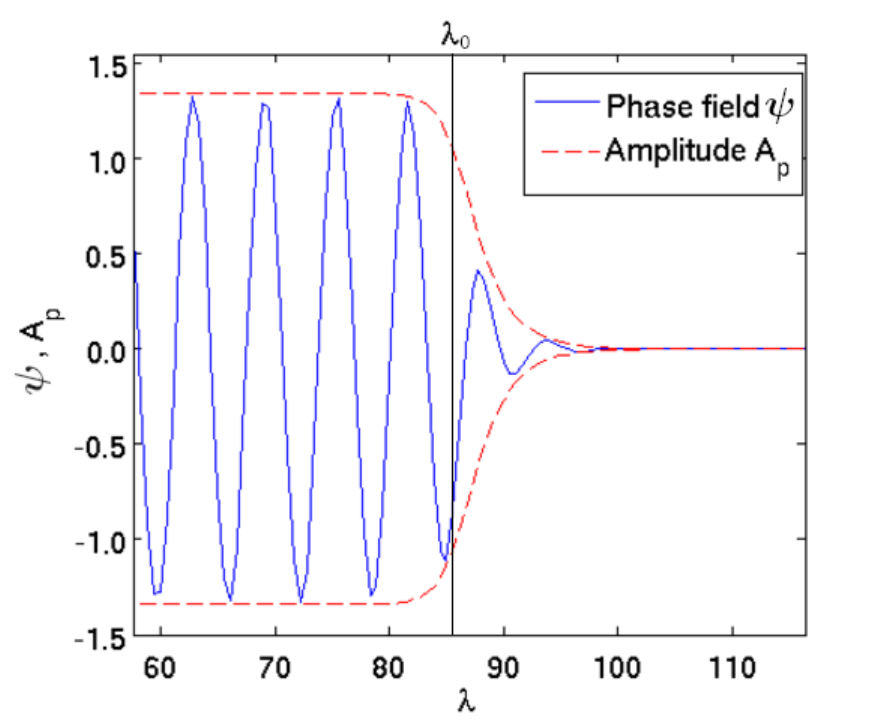}
\caption{Phase field order parameter profile $\psi$ along the normal direction $\lambda$ in a SmA-isotropic phase curved interface compared with the amplitude solution $A_p$ for $t = 2$. We have chosen $\epsilon_c = 0.675$. Further numerical details are given in Sec.~\ref{sec:results}. The function $A_p$ accurately captures the envelope of the field $\psi$.}
\label{fig:amp}
\end{figure}

If the smectic-isotropic interface is not planar, the amplitude $A$ will deviate from Eq.~(\ref{eq:amp-sol}). We expect, however, that for weakly curved interfaces, Eq.~(\ref{eq:amp-sol}) will be a good approximation when $z$ is replaced by the coordinate along the local normal to the interface. For example, Fig.~\ref{fig:amp} shows $A_p$ and the order parameter $\psi$ found from direct numerical solution of Eq.~(\ref{eq:pf-dynamics}), plotted along the local normal direction for the cyclide shown in Fig.~\ref{fig:ctorus} at time $t = 2$. Other than the location of the front, $z_{0}$, there are no adjustable parameters. The agreement between the two is excellent despite the fact that $\epsilon_{c} = 0.675$ is of order one. We also observed numerically that for values of $\epsilon$ up to 0.85 the front solution from Eq.~(\ref{eq:amp-sol}) still agrees with the interface obtained from the order parameter, even though it is no longer stationary.

Note that $A_{p}$ is not symmetric around $z_{0}$. In what follows, we will refer to the \lq\lq smectic-isotropic interface" as the locus of points of constant $A_{p}$, or, equivalently, of constant phase of $\psi$ in the front region. Appendix~\ref{sec:comp} discusses in detail how the location of the interface is obtained numerically from the order parameter $\psi$, and how the curvatures on the interface are computed. 


\subsection{Local equilibrium at curved front and kinetic law of motion}
\label{sec:GT}

Consider an idealized surface that corresponds to the smectic-isotropic interface, and let $\mathbf{p} = (s_1,s_2)$ be a point on the surface parametrized by $s_{1}$ and $s_{2}$. If $\lambda$ is the coordinate along the local normal to the surface ($\lambda = 0$ on the surface), the coordinates of a point $\mathbf{r}$ near the surface can be written as $\mathbf{r}(\lambda,s_1,s_2) = {\bf p}(s_1,s_2) + \lambda  {\bf n}\, (s_1,s_2)$, where $\mathbf{n}$ is the local normal at $\mathbf{p}$. The coordinates $s_1$ and $s_2$ are aligned with the principal directions, associated with the principal curvatures $c_1$ and $c_2$. We now seek solutions of Eq.~(\ref{eq:amp-gen}) of the form $A(\mathbf{r}) = A_{p} ( \lambda(\mathbf{r},t) )$.

We first compute the difference in chemical potential between a planar SmA-isotropic interface and a configuration with a weakly distorted interface, where the smectic layers remain parallel to the interface (perpendicular to the $\lambda$ direction). As previously noted, the phase $\phi$ of the amplitude is a constant near $\epsilon_c$, and the amplitude is a real quantity. The chemical potential $\mu$ in terms of the slowly varying amplitude $A$, is given by $\mu = \delta {\cal F}_{A}/\delta A$, and so 

\begin{equation}
    \mu \,=\,
        \epsilon A - 4\alpha q_0^2 \partial_z^2 A + \alpha \nabla^4_{xy} A
        - \frac{3}{4}\beta A^3 + \frac{5}{8}\gamma A^5.
\label{eq:mu-exp}
\end{equation}

The chemical potential $\mu_f$ for flat interface perpendicular to the $z$ direction can be directly obtained for a front $A$ aligned with $z$. In order to obtain the chemical potential $\mu_c$ associated with a curved interface, it is necessary to solve the corresponding amplitude equation. The scaling in $\epsilon$ for the coordinates transverse to the smectic-isotropic interface is $X = \epsilon^{1/4}x$ and $Y = \epsilon^{1/4} y$.  We assume that the same scaling applies to  $s_1$ and $s_2$. The induced scaling of the principal curvatures is $c_{1}, c_{2} \sim {\cal O}(\epsilon^{1/2})$, which follows from the fact that for small curvatures the mean curvature is half the trace of the Hessian matrix. The second derivative in the $z$ direction in Eq.~(\ref{eq:mu-exp}) generalizes to a second derivative in the normal direction $\lambda$. Additional contributions come from the curvatures, and are obtained by expanding the differential operators on local interface coordinates (Appendix~\ref{sec:lap-bel} details their expansion in terms of mean $H$ and Gaussian $G$ curvatures). We find, 

\begin{equation*}
\mu_c \;=\; \epsilon A - 4\alpha q_0^2 \bigg[\,\partial_\lambda-2H
          -(4H^2-2G)\,\lambda + 2H(3G-4H^2)\,\lambda^2\bigg]\,\partial_\lambda A
          -\frac{3}{4} \beta A^3 +\frac{5}{8} \gamma A^5\;.
\end{equation*}
For consistency, we have retained curvature terms below order $\epsilon^{7/4}$, the same order used in the derivation of Eq.~(\ref{eq:amp-gen}). Note that $(4H^2-2G)= c_1^2+c_2^2$ is known as the bending curvature. By multiplying both sides by $\partial_{\lambda} A_{p}$, integrating over $\lambda$, and subtracting the chemical potential for a planar surface (Appendix~\ref{sec:gibbs}) we find,

\begin{equation}
    \delta \mu \Delta A \;=\; 2H\sigma_h + (4H^2-2G)\sigma_b - 2H(3G-4H^2)\sigma_t.
    \label{eq:gt-par}
\end{equation}
This equation is the condition of local equilibrium, or the generalized Gibbs-Thomson equation in our model. The chemical potential difference between a curved and a planar surface $\delta \mu$ is given as a function of the surface curvatures, the discontinuity in amplitude between bulk smectic and isotropic phases, $\Delta A$, and three coefficients that depend explicitly on the one dimensional planar front solution $A_{p}$:

\begin{eqnarray}
  \nonumber
  \sigma_h
  &=&
      4\alpha q_0^2 \int^{\infty}_{-\infty}d\lambda\,(\partial_\lambda A_p)^2 
  \\[2mm]
  \nonumber
  \sigma_b
  &=&
      4\alpha q_0^2 \int^{\infty}_{-\infty}d\lambda\,(\partial_\lambda A_p)^2\lambda
  \\[2mm]
  \sigma_t
  &=&
      4\alpha q_0^2 \int^{\infty}_{-\infty}d\lambda\,(\partial_\lambda A_p)^2 \lambda^2
      \; .
    \label{eq:stress}
\end{eqnarray}
The first coefficient $\sigma_{h}$ is the standard surface tension coefficient that relates the change in chemical potential to the mean curvature of the surface. For weakly curved surfaces, this is the dominant term as it is inversely proportional to the radii of curvature. The second and third terms are of second and third order in the inverse radii of curvature, and describe deviations from the classical form of the Gibbs-Thomson equation. They represent interface bending ($\sigma_b$) and torsion ($\sigma_t$) contributions respectively, and are usually neglected. We retain all three terms in the expansion of the chemical potential in what follows because domains bounded by toroidal focal conics include regions in which the mean curvature vanishes, as well as regions of large curvature near the conic center. We will investigate numerically the accuracy of Eq.~(\ref{eq:gt-par}) in those regions. More generally, surface curvatures become large near regions of morphological singularities, and our result may extend the range of validity of the Gibbs-Thomson equation in the vicinity of the singularities. Finally, we stress that all three coefficients can be obtained from $A_{p}$ given in Eq.~(\ref{eq:amp-sol}), and therefore are completely determined by the parameters of the model, Eq.~(\ref{eq:pf-energy}).  Note in particular that $\sigma_{b} \neq 0$ because the solution $A_{p}$ is not symmetric around $z_{0}$.

A generalized Gibbs-Thomson equation similar to Eq.~(\ref{eq:gt-par}) has been previously given by Buff~\cite{buff1956curved} and Murphy~\cite{murphy1966thermodynamics} in the  context of curved fluid interfaces, albeit using different methods~\cite{markin1988definition}. The curvature terms in Eq.~(\ref{eq:gt-par}) coincide with theirs, except we have $2H(3G-4H^2) = -(c_1^3 + c_2^3)$ instead of $2HG$ alongside the interface torsion. Also, their curvature terms are associated with similarly defined coefficients $\sigma_h$, $\sigma_b$ and $\sigma_t$ (in fact, the terminology comes from the work of Murphy~\cite{murphy1966thermodynamics}).

A kinetic equation for the smectic-isotropic surface can be derived with a similar projection operation. The left hand side of Eq.~(\ref{eq:amp-gen}) is given by by $\partial_{t}A = \partial_{\lambda} (A_{p}) V_{n}$, where $V_{n}$ is the local normal velocity of the surface of constant $A_{p}$. The expansion of the right hand side of Eq.~(\ref{eq:amp-gen}) is the same as the right hand side of Eq.~(\ref{eq:mu-exp}). Multiplication by $\partial_{\lambda} A_{p}$ and integration over $\lambda$ (see Appendix~\ref{sec:gibbs}) gives the kinetic law of motion for the interface,

\begin{equation}
    V_n \;=\; -4\alpha q_0^2 \bigg\{2H + (4H^2-2G)\frac{\sigma_b}{\sigma_h} 
    - 2H(3G-4H^2)\frac{\sigma_t}{\sigma_h}\bigg\} .
    \label{eq:vel-par}
\end{equation}
The lowest order term is the classical law relating the normal velocity to the local mean curvature, while the remaining terms are the higher order  contributions (below $\epsilon^{7/4}$).  As is the case with Eq.~(\ref{eq:gt-par}), all coefficients are determined by the parameters of the model. 

The generalized Gibbs-Thomson equation~(\ref{eq:gt-par}), and the kinetic law, Eq.~(\ref{eq:vel-par}), have been derived under the assumption that the smectic layers are parallel to the smectic-isotropic interface. However, some of the configurations observed out of equilibrium in the experiments of Kim et al.~\cite{kim2016controlling} involve pyramidal structures in which smectic layers are exposed, so that they are aligned perpendicularly to the interface. In this case, for a planar interface the smectic layers are perpendicular to $z$ whereas the front normal is along $x$ (or $y$). The equation describing the planar front for this configuration is,

\begin{equation}
	-\epsilon A 
    - \alpha \,\partial_x^4 A + \frac{3}{4}\beta A^3 - \frac{5}{8}\gamma A^5 \;=\; 0.
    \label{eq:amp-perp}
\end{equation}
We cannot find an analytic solution for this front analogous to Eq.~(\ref{eq:amp-sol}), but it can be obtained numerically. For a weakly curved interface, a similar analysis to the previous case can be carried out, where the biharmonic from the amplitude equation~(\ref{eq:amp-gen}) is expanded when perturbations off coexistence are introduced in the weakly curved surface description (details given in Appendix~\ref{sec:lap-bel}). This calculation gives the change in chemical potential at a curved interface relative to planarity as, 

\begin{equation}
	\delta \mu \Delta A \;=\; \bigg[\frac{1}{2}\nabla^2_s H
    +2 H(H^2-G)\bigg] \frac{\sigma_h}{q_0^2}.
    \label{eq:gt-perp}
\end{equation}
The coefficient $\sigma_{h}$ is again given by Eq.~(\ref{eq:stress}), although in this case it must be computed approximately from the numerically determined solution of Eq.~(\ref{eq:amp-perp}). Importantly, however, the coefficient $\sigma_h/q_0^2$ is not a surface tension (energy per unit surface) due to the fact that the smectic layers are perpendicular to the interface in this configuration. In order to compute $\sigma_h$ for specific parameter values so as to carry out comparisons with the numerical solutions of the full phase field model (in Sec.~\ref{sec:results}), we have obtained a numerical solution of $A$ in Eq.~(\ref{eq:amp-perp}) through a finite difference relaxation method. For the parameter values of the model used ($q_0 = 1$, $\alpha = 1$, $\beta = 2$, $\epsilon = 0.675$ and $\gamma = 1$) we find that  $(\sigma_h)^\perp / (\sigma_h)^\parallel \approx 2.28$, which means that the effective tension for layers perpendicular to the interface is more than $100 \%$ larger than for layers parallel to the interface (see also Ref. \onlinecite{kim2016controlling}).

In analogy to the case with layers parallel to the interface, we can derive a kinetic law for the perpendicular interface. We find, 
\begin{equation}
	V_n \;=\; -4\alpha \bigg[\frac{1}{2}\nabla^2_s H
    +2H(H^2-G)\bigg] .
    \label{eq:vel-perp}    
\end{equation}
One remark about the derivation of Eqs.~(\ref{eq:gt-perp}) and~(\ref{eq:vel-perp}) is that integrals across the interface of the form $\sigma_{h2} = 4\alpha q_0^2\int d\lambda\,(\partial_\lambda^2A_p)(\partial_\lambda A_p)$ and $\sigma_{h3} = 4\alpha q_0^2\int d\lambda\,(\partial_\lambda^3A_p)(\partial_\lambda A_p)$ that appear in the derivation vanish in the limit of small $\epsilon$ since $\sigma_{h2}/ \sigma_{h}$ and $\sigma_{h3} / \sigma_{h}$ scale as $\epsilon^{1/4}$ and $\epsilon^{1/2}$ respectively. The kinetic equation~(\ref{eq:vel-perp}) that results has a form similar to that of a Willmore flow~\cite{willmore1996riemannian}, although it differs by a factor of 1/2 in the surface Laplacian. Similar kinetic laws (also called fourth order flows) in which the biharmonic operator plays a role in the dynamics~\cite{du2004phase,burger2008willmore} have been examined in connection with the biharmonic heat equation and the Willmore flow~\cite{koch2012geometric}.


\section{Numerical study of toroidal focal conic instabilities}
\label{sec:results}

We use the phase field model given by Eqs.~(\ref{eq:pf-energy}) and~(\ref{eq:pf-dynamics}) to study the evolution of a single focal conic domain of a smectic phase in contact with an isotropic phase. The computational cell is a three dimensional cubic mesh of size $512^{3}$ or $1024^{3}$. Boundary conditions of the computational domain are zero normal derivatives of $\psi$, and zero normal derivative of the Laplacian of $\psi$. Focal conic domains, when present, are compatible with these boundary conditions, since they favor parallel alignment of the molecules with respect to the boundaries. Unless otherwise noted, we use $\alpha = 1$, $\beta = 2$ and $\gamma =1$ in our calculations. These parameters yield a coexistence value of $\epsilon_{c} = 0.675$. We also use $q_{0} = 1$ as the reference wavenumber. The focal conic configuration used for initial conditions (e.g., Fig.~\ref{fig:ctorus}) is defined by $\psi(\lambda) = A \,\textrm{cos}(q_0 \lambda)$ in the smectic, where $\lambda$ is the normal direction, $q_0 = 1$, and then amplitude $A$ is given by Eq.~(\ref{eq:amp-par}). This phase is in contact with an isotropic phase $\psi = 0$.

Equation~(\ref{eq:pf-dynamics}) is solved numerically by a pseudo-spectral method, in which gradient terms are computed in Fourier space and nonlinear terms in real space. Space discretization, based on 16 points per wavelength, is $\Delta x  = 2 \pi /(16 q_{0})$. Integration in time is of second order with a Crank-Nicholson algorithm for the linear part of the equation, and a second order Adams-Bashforth method for the nonlinear terms. The time step used is $\Delta t = 5 \cdot 10^{-4}$. We have developed a custom C++ code based on the parallel FFTW library and the standard MPI passing interface for parallelization. In order to accommodate the stated boundary conditions, we use the Discrete Cosine Transform. Further details on the computational method, tracking of the the smectic-isotropic surface, and calculation of the interfacial curvatures can be found in Appendix~\ref{sec:comp}.

As discussed earlier, the value that we choose for $\epsilon_c$ allows for a reasonably large region of coexistence. This is advantageous from a numerical standpoint, as for small $\epsilon_c$ it is a challenging task to maintain coexistence in three dimensions. At the same time, since the interface equations were derived for small values of $\epsilon$, we had to perform a numerical check to confirm that we were still within the  limit of validity of the asymptotic analysis. We observed that while the  solution $A_p$ from Eq.~(\ref{eq:amp-sol}) was derived for small $\epsilon$,  it still accurately describes the envelope for $\epsilon_c = 0.675$, as  observed along the curved interface of a focal conic in Fig.~\ref{fig:amp}. Even when displacing the system from coexistence, the solution $A_p$ remains an approximate description of the interface, up to $\epsilon \sim 0.85$. The validity of the interface equations for this value of $\epsilon_c$ is further confirmed by our numerical results for the interface velocity, as will be presented in this section.


\subsection{Stationary Clifford torus}

In order to verify the accuracy of the numerical scheme, we first consider a toroidal configuration at coexistence $\epsilon = \epsilon_c$, and examine smectic planes bent in the shape of a focal conic. Friedel~\cite{friedel1922etats} was the first to associate focal conic domains with Dupin cyclides, arguing that smectic molecular layers would bend in this geometrical fashion while remaining parallel to the interface. Later, these cyclides were also shown to be stable configurations of a SmA via energy minimization of the Oseen-Frank energy given by Eq.~(\ref{eq:of-energy})~\cite{bragg1934liquid,frank1958liquid,schief2005nonlinear}. If the layer spacing of the smectic in equilibrium is assumed to remain approximately constant, and given that the term proportional to the Gaussian curvature is a null Lagrangian, minimization of Eq.~(\ref{eq:of-energy}) reduces to the minimization of $\int d{\bf x}(K/2) H^2$, where $K$ is the splay elastic modulus. This is the classical Willmore problem. Surfaces that minimize this energy are Willmore surfaces, which include minimal surfaces, spheres, and Dupin cyclides (in particular, the axially symmetric Clifford torus), and are obtained by an evolution that follows the Willmore flow~\cite{willmore1996riemannian}. 

\begin{figure}[h]
	\centering
    \begin{subfigure}[b]{0.33\textwidth}
        \includegraphics[width=\textwidth]{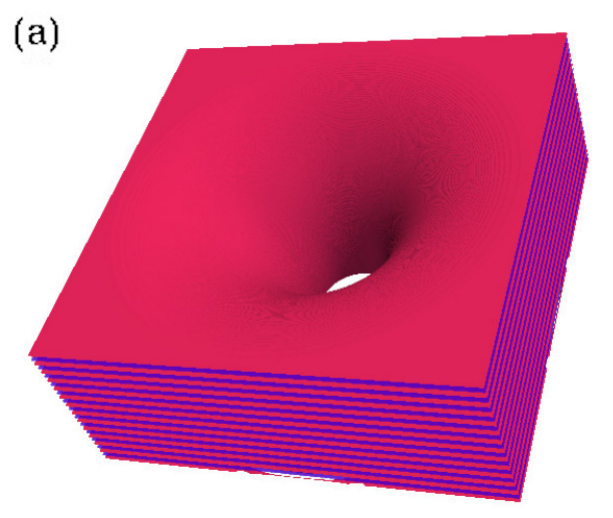}
    \end{subfigure}
    \hspace{15mm}
    \begin{subfigure}[b]{0.33\textwidth}
        \includegraphics[width=\textwidth]{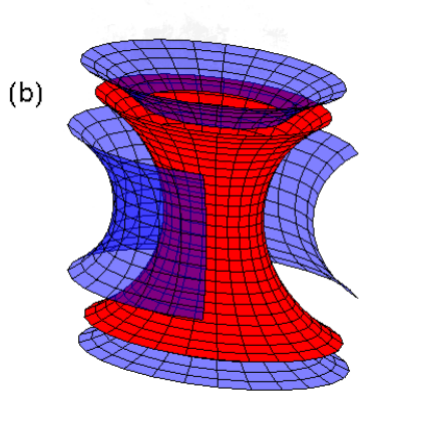}
    \end{subfigure}
    \caption{(a) Clifford torus configuration as represented by the phase 
    field. (b) For reference, we show internal segments for a family 
    of Clifford tori.}
	\label{fig:ctorus}
\end{figure}

\begin{figure}[h]
	\centering
    \begin{subfigure}[b]{0.44\textwidth}
    \includegraphics[width=\textwidth]{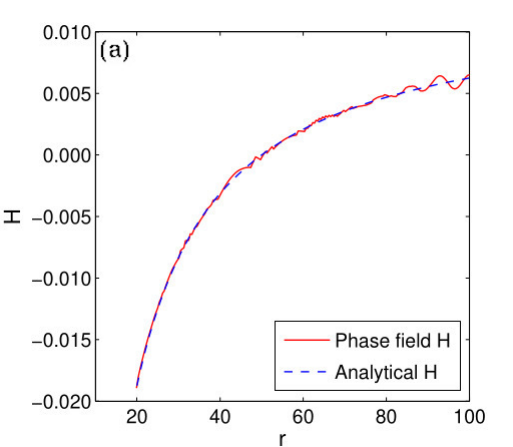}
    \end{subfigure}
    \begin{subfigure}[b]{0.44\textwidth}
        \hspace*{3mm}
    \includegraphics[width=\textwidth]{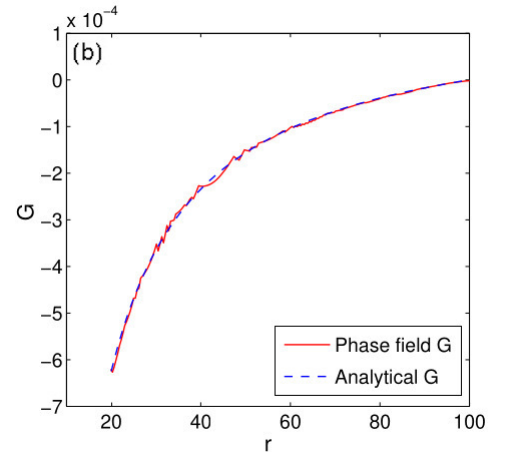}
    \end{subfigure}
    \caption{Stationary values of the (a) mean and (b) Gaussian 
    curvatures computed for the SmA surface from the phase field in Fig.~\ref{fig:ctorus}.
    They are plotted along the radial direction ($r=0$ at the center of the torus),
    and compared with the analytic curvatures of a Clifford 
    torus. We use $N = 512^{3}$ and coexistence parameters, with
    $\alpha = 1$, $\beta = 2$, $\gamma = 1$ and $\epsilon = 0.675$.}
	\label{fig:ctcurv}
\end{figure}

We have verified that stationary solutions of the phase field model agree with this result. We consider an initial condition with layers bent in a cyclide configuration, as a half-torus, in which there is no self-intersection of layers; hence, we have a disk of isotropic phase inside the torus and in contact with the substrate. We then compute the evolution of this configuration by integrating Eq.~(\ref{eq:pf-dynamics}). The evolution leads to the stationary Clifford torus shown in Fig.~\ref{fig:ctorus}. Every cross section along the radial direction will display two sections of the torus. We show our numerical results in Fig.~\ref{fig:ctcurv} for both mean and Gaussian curvatures of a cross section of the surface. They agree very well with the curvatures obtained from an analytic Clifford torus of the same size.  

The circular arrangement of the planes seen from a cross section in the radial direction is known as a target pattern in the phase field literature~\cite{korzinov1993origin}, such that we can observe two quarter circle targets in a cross section, one on each side of the center hole. The target pattern is a stationary solution of Eq.~(\ref{eq:pf-dynamics}) in two dimensions. This can be seen by writing Eq.~(\ref{eq:amp-gen}) in polar coordinates, with $r$ the radial coordinate and $r = 0$ at the center of the target. The solution for $r \gg 1$ is $A(r) = \sqrt{1 - 1/r^2}\,A_s$, where $A_s$ is the solution for the polynomial part of the amplitude equation given by Eq.~(\ref{eq:amp-sma}). Since the Clifford torus is an axially symmetric cyclide, this observation about the target patterns implies that such a torus should also be a solution for Eq.~(\ref{eq:pf-dynamics}), as verified in Fig.~\ref{fig:ctorus}.

\subsection{Evolution of focal conic domains at coexistence}

\begin{figure}[h]
	\centering
    \begin{subfigure}[b]{0.40\textwidth}
        \includegraphics[width=\textwidth]{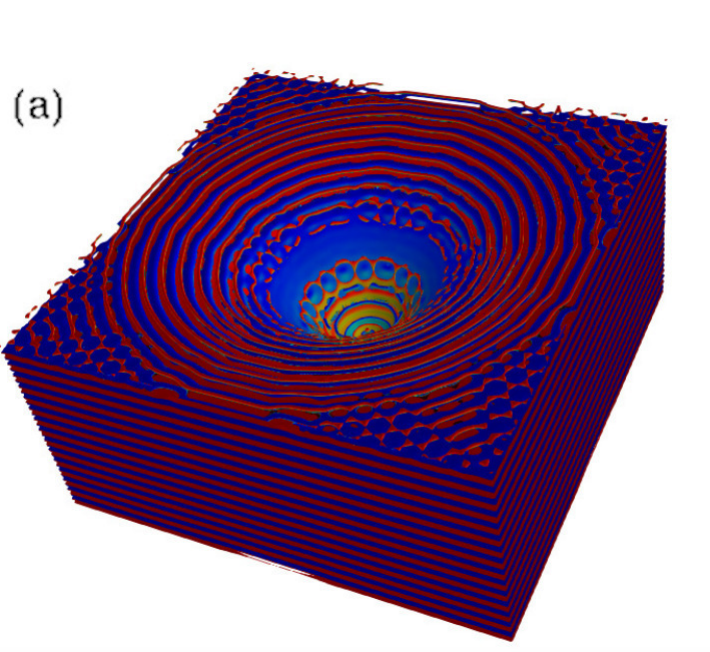}
    \end{subfigure}
    \hspace{5mm}
    \begin{subfigure}[b]{0.4\textwidth}
        \includegraphics[width=\textwidth]{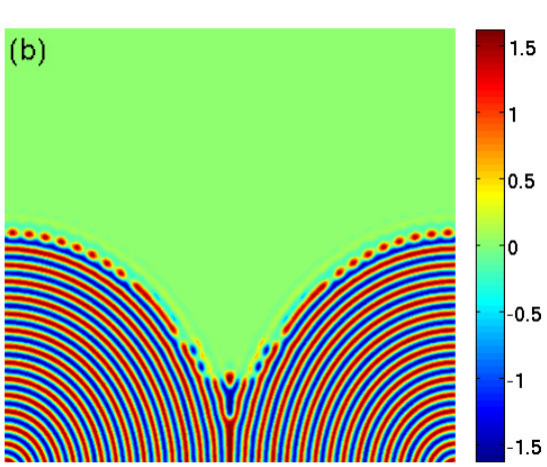}
    \end{subfigure}
    \caption{(a) Three dimensional phase field and
        (b) cross section for a focal conic
        which is unstable at its core, extracted from time $t = 150$.
        Parameters are set within the coexistence region ($\alpha = 1$,
        $\beta = 2$, $\gamma = 1$ and $\epsilon = \epsilon_c = 0.675$).}
    \label{fig:deepfc}
\end{figure}

We consider a focal conic at coexistence involving a macroscopic cusp where smectic layers self intersect. This initial configuration is no longer stationary, and the evolution of the order parameter is shown in Fig.~\ref{fig:deepfc}. Near the cusp, where the mean curvature is negative, a small smectic region nucleates, whereas in the outer region of positive mean curvature, smectic layers near the interface evaporate. A stationary configuration is reached which is shown in the figure. Smectic condensation at the cusp like depression is also observed by experiments, where material transfers along the interface owing to the variation of the local vapor pressure at the interface~\cite{kim2016controlling}.

\begin{figure}[h]
	\centering
    \begin{subfigure}[b]{0.44\textwidth}
    \includegraphics[width=\textwidth]{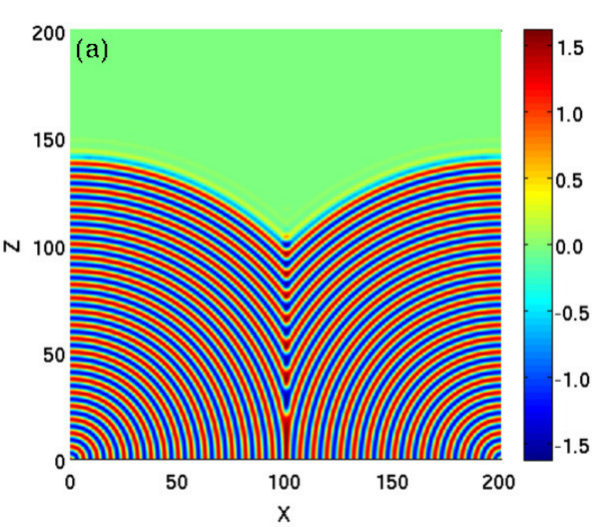}
    \end{subfigure}
    \hspace{5mm}
    \begin{subfigure}[b]{0.43\textwidth}
    \includegraphics[width=\textwidth]{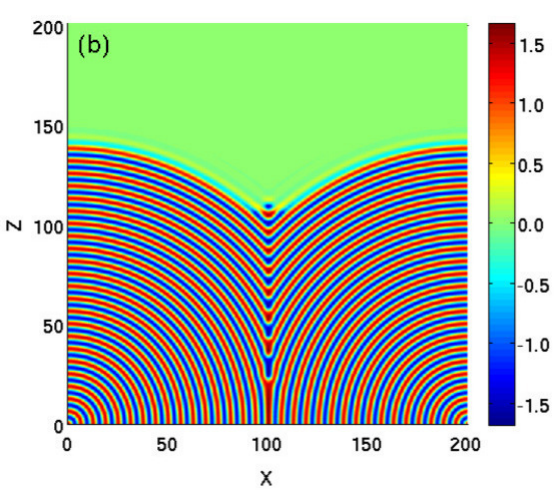}
    \end{subfigure}
    \caption{Cross section of the phase field order parameter
        representation for a TFCD at coexistence ($\alpha = 1$,
        $\beta = 2$, $\gamma = 1$ and $\epsilon = 0.675$).
        (a) Numerical solution for the starting stage. (b) Later stage,
        time $t = 50$, we see some deposition at the core of the defect.}
	\label{fig:fc512}
\end{figure}

Figure~\ref{fig:fc512} shows results for a similar initial configuration, but with a larger number of smectic layers. This configuration is closer to the focal conics observed in SmA films, and illustrates the instability of the layer cusps deep inside the smectic domain. Curvatures are smaller in magnitude when compared to the previous case, in particular close to the singularity, which slows down the dynamics. We still observe some condensation at the core under coexistence, but no evaporation is seen near the boundaries. This chevron pattern has also been observed in phase field models of low angle grain boundaries~\cite{re:tsori00}.

\subsection{Evolution of focal conic domains away from coexistence}

\begin{figure}[h]
 	\centering
    \begin{subfigure}[b]{0.4\textwidth}
    \includegraphics[width=\textwidth]{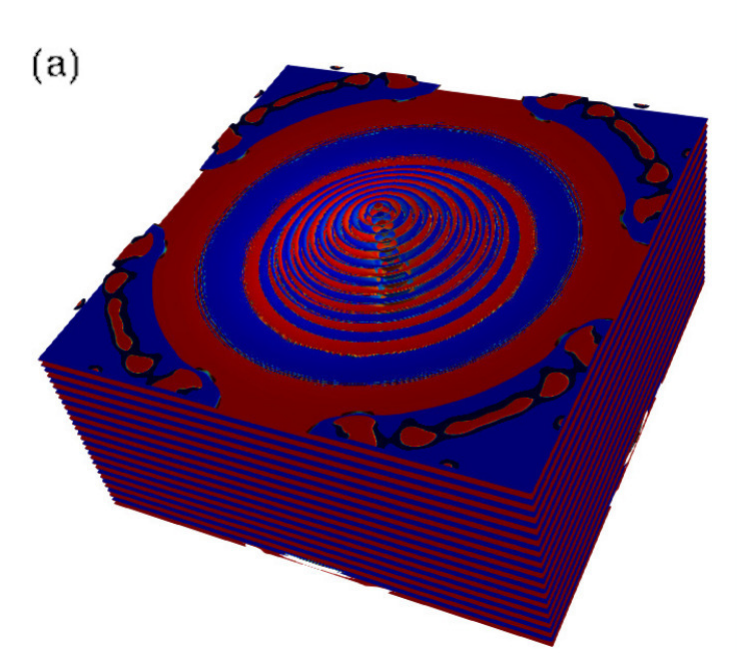}
    \end{subfigure}
    \hspace{5mm}
    \begin{subfigure}[b]{0.4\textwidth}
    \includegraphics[width=\textwidth]{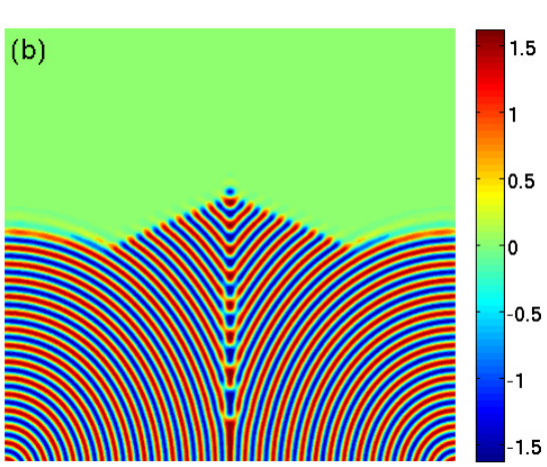}
    \end{subfigure}
    \caption{(a) Three dimensional phase field and (b) cross section
        for a conical pyramid that appears due to the localized
        evaporation of smectic layers around the edges, extracted from time
        $t = 50$. Parameters are set such that the isotropic phase is
        thermodynamically favored ($\alpha = 1$, $\beta = 2.0$,
        $\gamma = 1.0$ and $\epsilon = 0.8$).}
    \label{fig:cpyramid}
\end{figure}

We next study the evolution of a toroidal focal conic initial condition away from coexistence.  We take $\epsilon > \epsilon_c$, which corresponds to a thermal treatment in which the isotropic phase has lower free energy than the smectic. The initial configuration is similar to one considered in Fig.~\ref{fig:fc512}, but with more smectic layers. We observe that smectic layers in the outer region evaporate, leading to a conical pyramid in the center, as shown in Fig.~\ref{fig:cpyramid}. The evaporation of each layer stops once the layer border aligns with the one above, creating an interface of stacked layers. The pyramid has positive Gaussian curvature, in contrast to the initial layers of negative Gaussian curvature. Similar pyramidal morphologies are observed experimentally~\cite{kim2016controlling}.

During the evaporation of the smectic film, we compare the numerically computed interface normal velocity, given by $V_n = \partial_t \psi / |\nabla\psi|$ with the asymptotic predictions of Eqs.~(\ref{eq:vel-par}) and~(\ref{eq:vel-perp}). We consider first the case of smectic layers parallel to the interface, with velocity described by Eq.~(\ref{eq:amp-par}). The initial configuration adopted is the same as the one used to generate Fig.~\ref{fig:fc512}. We take $\epsilon = 0.75$, and all numerical data shown corresponds to the initial stages of evolution ($t = 5$) so that the SmA layers remain parallel to the interface across the entire surface outside a small neighborhood around the cusp. The values of the coefficients $\sigma_h$, $\sigma_b$ and $\sigma_t$ used are given in Eq.~(\ref{eq:stress}) with $A_{p}$ defined in Eq.~(\ref{eq:amp-sol}). Local mean and Gaussian curvatures are directly obtained from the evolving phase field as discussed in Appendix~\ref{sec:comp}.  Figure~\ref{fig:velpar} shows the normal velocity computed from the full phase field model, the normal velocity predicted by Eq.~(\ref{eq:vel-par}), and the normal velocity that follows from mean curvature motion alone (i.e., with $\sigma_{b}=\sigma_{t} = 0$). The system size is $N = 1024^3$ so that $0 < x < 401 $. The interface singularity is located at $x \approx 200$ in the figure.  While there is good agreement among all three results away from the center, differences appear in the high curvature region towards the center of the domain.  Specifically, motion driven by mean curvature alone near the focal conic center deviates from the computed interface velocity, including its sign. On the other hand, the normal velocity predicted by the higher-order velocity equation agrees with the numerical value until very close to the center of the focal conic. We note that there are no adjustable parameters in the results shown in Fig.~\ref{fig:velpar}, except for a uniform velocity shift owing to the lower energy of the isotropic phase, as $\epsilon > \epsilon_{c}$.  We observe that the region in which mean curvature driven growth deviates from the full numerical calculation is rather small. We estimate that the radius of this region would be on the order of 30 nm in the experiments of Ref. \onlinecite{kim2016controlling}, and hence below the resolution of optical detectors. Nevertheless, our calculation is consistent with the experimental observation that pyramids form due to smectic layer evaporation away from the focal conic center, not nucleation of new smectic layers at the center. 

\begin{figure}[h]
    \centering
    \includegraphics[width=0.5\textwidth]{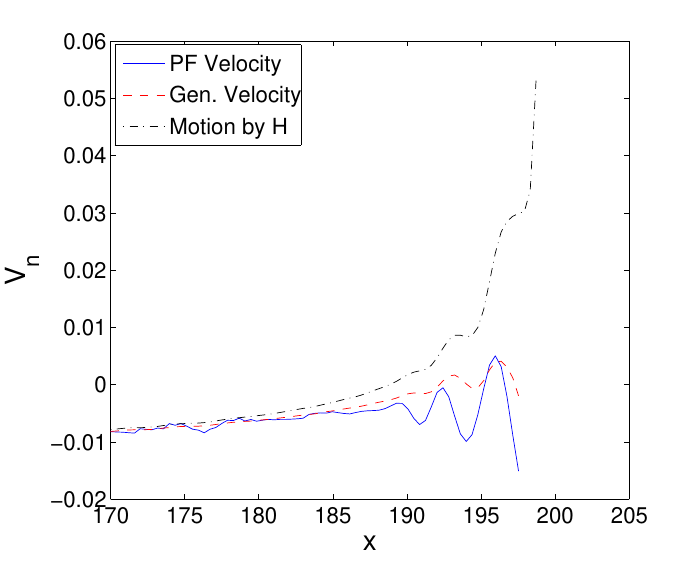}
    \caption{Local normal velocity of SmA-isotropic interface, extracted from
        a focal conic under sintering ($\epsilon = 0.75 > \epsilon_c$).
        The numerically determined surface velocity is plotted against the 
        generalized velocity prediction for planes parallel to the interface,
        and compared to the classical prediction of mean curvature driven motion.
        $N = 1024^3$, defect core at $x \approx 200$.}
 	\label{fig:velpar}
\end{figure}

\begin{figure}[h]
 	\centering
    \begin{subfigure}[b]{0.45\textwidth}
    \includegraphics[width=\textwidth]{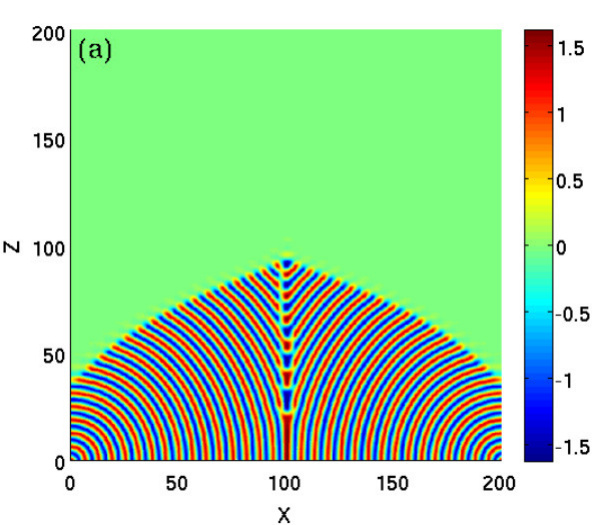}
    \end{subfigure}
    \hspace{5mm}
    \begin{subfigure}[b]{0.45\textwidth}
    \includegraphics[width=\textwidth]{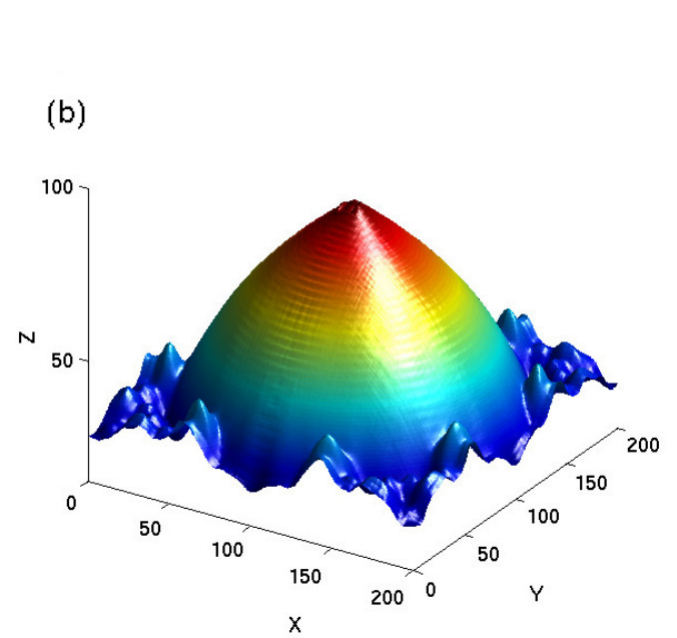}
    \end{subfigure}
    \caption{(a) Cross section and (b) interface location
        for a pyramidal morphology obtained from a focal conic under
        thermal sintering, extracted from time $t = 200$. Coloring of the
        interface location indicates the height $z$. Initial condition was
        composed of a focal conic with layers reaching almost the top ($z = 200$)
        of the domain, using $N = 512^3$. Parameters are such such that the
        isotropic phase is thermodynamically favored ($\alpha = 1$, $\beta = 2.0$,
        $\gamma = 1.0$ and $\epsilon = 0.8$).}
 	\label{fig:latecpyr}
\end{figure}

\begin{figure}[h]
 	\centering
    \begin{subfigure}[b]{0.45\textwidth}
    \includegraphics[width=\textwidth]{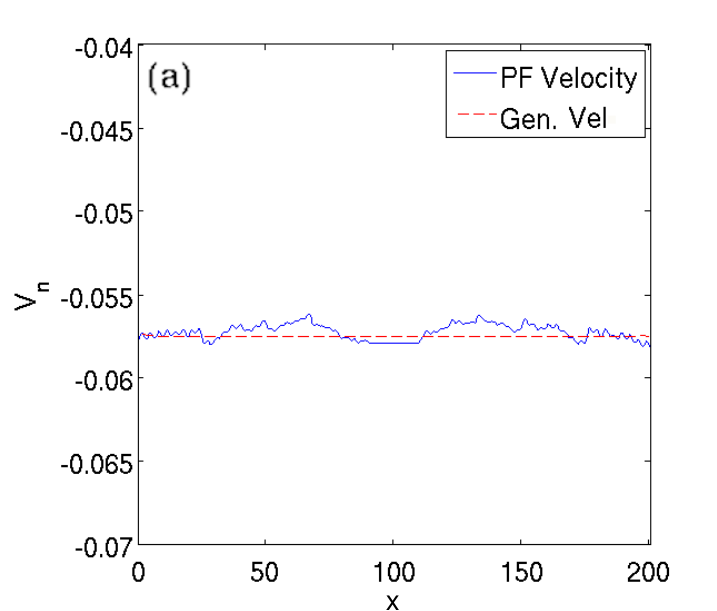}
    \end{subfigure}
    \hspace{5mm}
    \begin{subfigure}[b]{0.45\textwidth}
    \includegraphics[width=\textwidth]{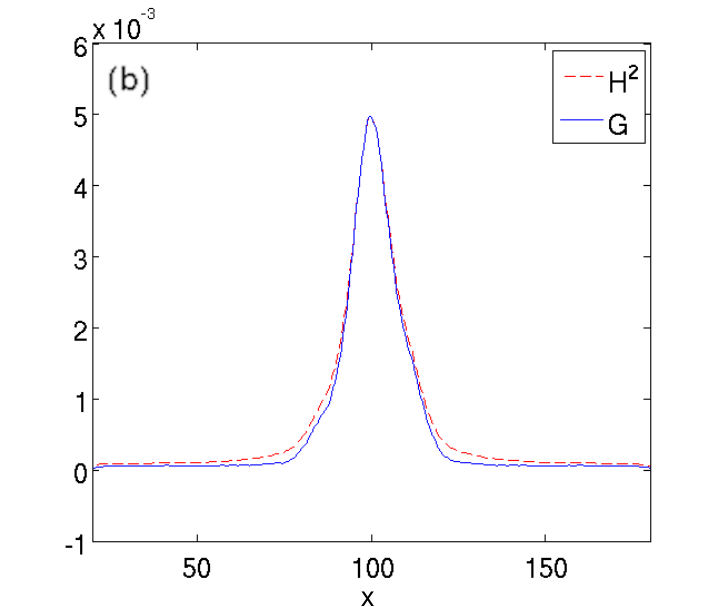}
    \end{subfigure}
    \caption{Interface velocity and curvature comparison for the
        pyramid with $\epsilon = 0.8$ shown in Fig.~\ref{fig:latecpyr}.
        (a) The numerically determined surface velocity is plotted against
        the generalized kinetic law for planes perpendicular to the interface.
        (b) mean curvature squared $H^2$ approximately matches the
        Gaussian curvature $G$ for this morphology.}
 	\label{fig:infocpyr}
\end{figure}

As mentioned previously, the results shown in Fig.~\ref{fig:velpar} were taken early in the evolution, so that the pyramidal structure was just beginning to form.  As the pyramidal structure grows to macroscopic size, as in Fig.~\ref{fig:cpyramid}, the smectic planes in the pyramid are perpendicular, not parallel, to the smectic-air interface. This agrees with the observed morphological reconstruction of smectic films during thermal sintering~\cite{kim2016controlling}. As a consequence, the local normal velocity in this case should be given by Eq.~(\ref{eq:vel-perp}). Consider a large pyramidal structure, shown in Fig.~\ref{fig:latecpyr}, taken from a calculation with $N = 512^3$, $\epsilon = 0.8$ and after a fairly long time of $t = 200$. As observed, the pyramidal surface is smooth enough for the curvatures to be computed without issues, and the corresponding interfacial velocity is shown in Fig.~\ref{fig:infocpyr} (left). We find that the normal velocity is approximately constant and slightly negative over the entire pyramid, meaning that the structure shown is uniformly evaporating, albeit slowly. The curvatures of the moving interface are shown in Fig.~\ref{fig:infocpyr} (right). The mean curvature squared $H^2$ is almost identical to the  Gaussian curvature $G$, which, given the interfacial kinetic equation Eq.~(\ref{eq:vel-perp}), accounts for the small and almost constant normal velocity over the entire pyramid interface. The constant rate of evaporation is due to the difference in bulk energy between the two phases when $\epsilon > \epsilon_c$, and does not depend on local curvatures.

\begin{figure}[h]
 	\centering
    \begin{subfigure}[b]{0.44\textwidth}
    \includegraphics[width=\textwidth]{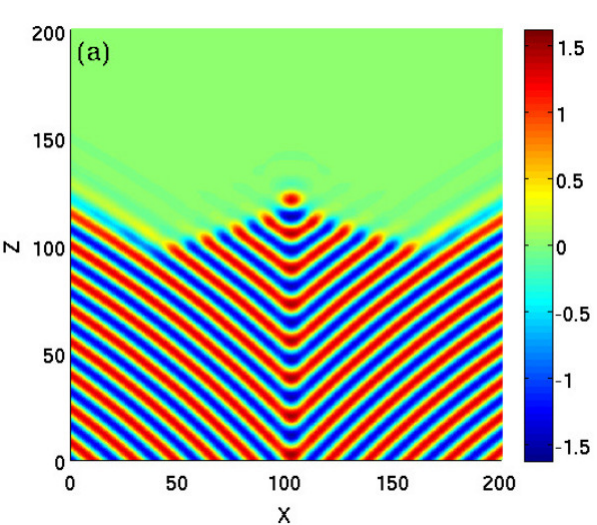}
    \end{subfigure}
    \hspace{5mm}
    \begin{subfigure}[b]{0.44\textwidth}
    \includegraphics[width=\textwidth]{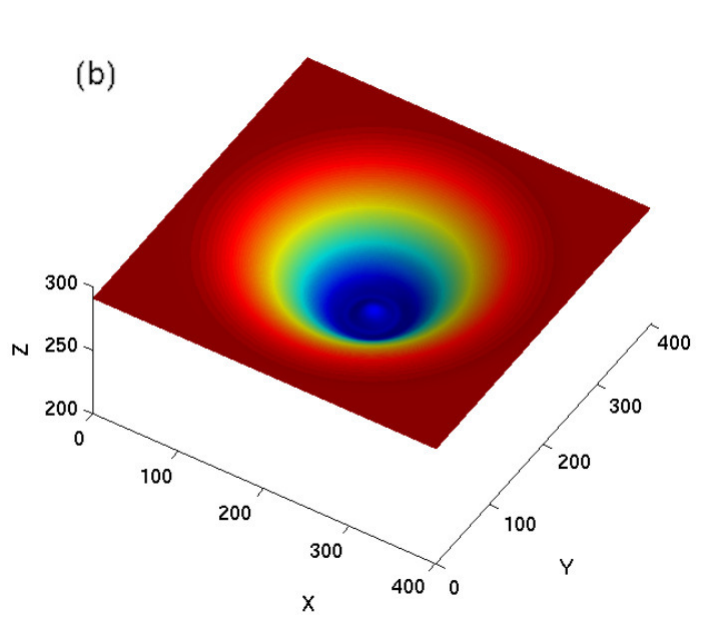}
    \end{subfigure}
    \caption{(a) Expanded cross section blow up and (b) interface location right
        for a focal conic during thermal sintering, showing a 
        pyramid of appreciable size, extracted from time $t = 50$.
        We used $N=1024^3$, for which the numerical solution reveals the pyramidal
        structure being formed at the core around layers that remain
        parallel to the interface. Parameters are such such that the
        isotropic phase is thermodynamically favored ($\alpha = 1$, $\beta = 2.0$,
        $\gamma = 1.0$ and $\epsilon = 0.8$).}
 	\label{fig:mixed}
\end{figure}

\begin{figure}[h]
 	\centering
    \begin{subfigure}[b]{0.45\textwidth}
    \includegraphics[width=\textwidth]{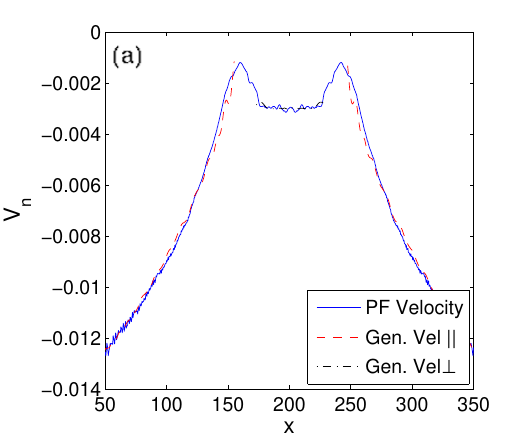}
    \end{subfigure}
    \hspace{5mm}
    \begin{subfigure}[b]{0.45\textwidth}
    \includegraphics[width=\textwidth,height=0.84\textwidth]{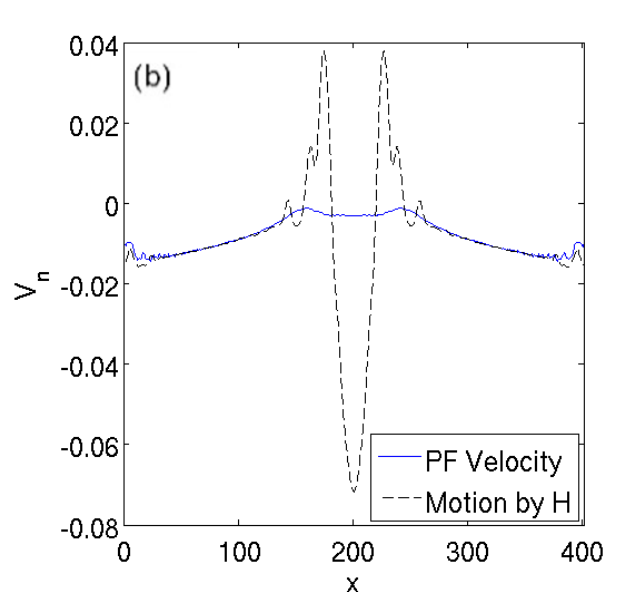}
    \end{subfigure}
    \caption{Interface velocity for a middle cross-section. 
        (a) The numerically determined surface velocity is plotted 
        against the two generalized kinetic laws, one for each region.
        (b) The evolution by mean curvature velocity prediction strongly
        diverges in the central pyramidal region.}
 	\label{fig:velmixed}
\end{figure}

We conclude by presenting numerical results for a larger system ($N = 1024^{3}$), with $\epsilon = 0.8$, so that we can examine the different interface orientations within a single numerical solution. The initial configuration is a focal conic domain. As the configuration evolves, smectic layers away from the middle (and parallel to the interface) evaporate while a pyramid (with layers perpendicular to the interface) forms at the center.  The transient morphology obtained at $t = 50$ is shown in Fig.~\ref{fig:mixed}. The local normal velocity in the outer region is given by Eq.~(\ref{eq:vel-par}), whereas the inner region local normal velocity is given by Eq.~(\ref{eq:vel-perp}). As was the case in the experiments of Ref. \onlinecite{kim2016controlling}, the conical pyramid forms due to curvature induced evaporation of layers in the outer region, whereas evaporation is essentially negligible in the pyramidal region owing to the balance of mean and Gaussian curvatures. Our numerically obtained normal velocities for this interface are shown in Fig.~\ref{fig:velmixed}(left). As before, there is a constant background shift of both curves arising from the the constant energy difference between the bulk phases, but there are otherwise no adjustable parameters. The agreement between the numerical solution and the predictions of the asymptotic analysis is excellent.

Finally, we show in Fig.~\ref{fig:velmixed}(right) the interfacial normal velocity that would result from mean curvature driven growth alone. The agreement with the numerical result is quite good in the outer region of small curvature, where the effects of bending and torsion are negligible. Near the center, however, mean curvature driven growth fails to describe the numerical results.


\section{Conclusions and discussion}

In this work we have introduced a model for a smectic-isotropic system from which we have derived a generalized Gibbs-Thomson and interfacial motion equations, revealing the role of the Gaussian curvature and the orientation of a modulated phase on local equilibrium thermodynamics and kinetics of the interface. The computational challenges of tracking a complex and moving smectic-isotropic phase boundary have been addressed by using a phase field model. We have presented an asymptotic analysis of the solutions of the model, valid near smectic-isotropic coexistence, and for weakly curved interfaces. Through this analysis we obtain a dynamical equation for the amplitude that modulates the periodic smectic layering. This procedure allows us to obtain physical insights about local equilibrium thermodynamics and evaporation-condensation dynamics of the smectic-isotropic interface without dealing with the oscillatory nature of the layering description. 

The work is directly motivated by recent experiments in the sintering of toroidal focal conic domains in thin films of smectic liquid crystals~\cite{kim2016controlling,kim2018curvatures}, which show novel morphologies, including conical pyramids and concentric rings. By both simulating the sintering of focal conics and comparing the results to an asympotic analysis of the governing equations, we reproduce the evaporation process that takes place in the experiments, while clarifying the limitations of classical interface equations. Our results portray how focal conics evolve to conical pyramids through evaporation and condensation of the smectic layers, as observed in the experiments. The analysis also shows that when smectic planes are parallel to the interface, three surface energy coefficients are necessary to describe local equilibrium thermodynamics and kinetics to the order of approximation considered. These coefficients can be computed analytically within the model, from Eq.~{\ref{eq:stress}}. For the case of planes perpendicular to the interface, the chemical potential at a curved interface is not proportional to the local mean curvature, but rather a Willmore type problem emerges.

Our findings expand the range of understanding and control of micropatterning of smectic films, as templates for superhydrophobic surfaces~\cite{kim2009fabrication}, guides for colloidal dispersion~\cite{milette2012reversible,pratibha2010colloidal} and soft lithography~\cite{yoon2007internal,kim2010self}. More broadly, the present results can guide future experiments in other modulated phases such as block copolymers. Our generalized theory should also benefit research in biomembranes, which have already reported connections between the Gaussian curvature and protein binding~\cite{re:elliott15}, as well as work on nucleation and growth on curved surfaces~\cite{re:gomez15}. 

We mention finally that our analysis focuses on the smectic-isotropic interface, whereas the experiments in thin films concern a smectic-air interface instead. Therefore our analysis does not contain any hydrodynamic stresses at the smectic-air boundary, or any resulting flows. Although velocity fields were not measured in the experiments, and the results were interpreted in terms of the same evaporation-condensation mechanisms that we have examined here, the excess energies that introduce corrections to the Gibbs-Thomson equation will also lead to normal stresses at the boundary. Work that includes these stresses is currently in progress.


\acknowledgments

We are indebted to Oleg Lavrentovich for many stimulating discussions. This
research has been supported by the Minnesota Supercomputing Institute, and by
the Extreme Science and Engineering Discovery Environment (XSEDE)
\cite{towns2014xsede}, which is supported by the National Science Foundation
grant number ACI-1548562.  E. Vitral would like to thank the continuous support
from the Aerospace Engineering and Mechanics Department - University of
Minnesota during the entire length of this work. JV is supported by the National
Science Foundation, contract DMR-1838977.


\newpage

\appendix

\section{Amplitude equation}
\label{sec:amp-eqn}

The phase field order parameter introduced in Sec.~\ref{sec:model} is driven
by energy minimization, with the following dynamical equation
\begin{eqnarray}
    \partial_t \psi \;=\; -\epsilon \,\psi 
    - \alpha \, (q_0^2 + \nabla^2)^2 \psi + \beta \,\psi^3 - \gamma \,\psi^5.
  \label{eq:pf-dynamics2}
\end{eqnarray}
Our goal is to derive an amplitude equation~\cite{manneville1995dissipative,cross1993pattern,burke2006localized} describing the motion of the envelope that describes the SmA-isotropic front without the oscillatory behavior of the phase field. We perform this analysis for small positive values of $\epsilon$,  $\epsilon \ll 1$ such that the amplitude of the order parameter is also small. Assuming the SmA layers are perpendicular to the $z$ direction, the solution representing this phase is approximately $\psi({\bf x}, t) \approx \frac{1}{2}(A e^{iq_0z} + c.c.)$. Space and time can be separated in fast and slow scales, where the fast variables are $\{x, y, z, t\}$, and the slow variables are $\{X, Y, Z, T\}$. If we consider this amplitude to be slowly modulated along the perpendicular direction to the layers, we can set a distinction between the fast varying carrier exp$(iq_0z)$, and the slowly varying the amplitude $A(X, Y, Z, T)$.

The slow variables scaling can be obtained by introducing small perturbations in the different directions. Although the energy is rotationally invariant, perturbations in $x$, $y$ and $z$ will scale differently. For instance, take perturbations in z, $\psi = \psi(x,y,z+\delta_z,t)$, and linearize Eq~{\ref{eq:pf-dynamics2}}

\begin{equation}
    \partial_t \psi \,=\, \big[ -\epsilon - \alpha( 2q_0 \delta_z + \delta_z^2)^2 \big]\psi
    \,=\, \big[ -\epsilon - \alpha( 4q_0^2 \delta_z^2 + 4q_0 \delta_z^3+ \delta_z^4) \big]\psi \; .
\label{eq:dist-z}
\end{equation}
Now, compare it to perturbations along $x$ (or $y$), $\psi = \psi(x+\delta_x,y,z,t)$:

\begin{equation}
    \partial_t \psi \,=\, \big( -\epsilon - \alpha \, \delta_x^4 \big)\psi
\label{eq:dist-x}
\end{equation}
From Eqs.~{\ref{eq:dist-z}} and {\ref{eq:dist-x}}, we observe that the consistency condition between the lowest order terms acting on the slowly modulated envelope should be

\begin{equation*}
\partial_t A \;\sim\; \epsilon A \;\sim\;
\partial_z^2A \;\sim\; \partial_x^4 A \;\sim\; \partial_y^4 A \;.
\end{equation*}
Hence, the slow variables scale as

\begin{equation}
    X = \epsilon^{1/4}x, \;\;\; Y = \epsilon^{1/4}y, \;\;\;
    Z = \epsilon^{1/2} z, \;\;\; T = \epsilon t \;. 
\end{equation}
Note that $\beta \sim \epsilon^{1/2}$, since at the coexistence point $\epsilon_c = 27\beta^2/160\gamma$. Also, one can show that both $\psi = 0$ and the non-trivial solution are stable for $\epsilon > 0$ up to the turning point $\epsilon_{tp} = 9\beta^2/40$.  For larger $\epsilon$ only the trivial solutions exists and is stable. For small values of $\epsilon$ these two points become very close, and they are also within the range of small perturbations from the bifurcation point $\epsilon = 0$.

From the proposed scaling and the chain rule, the derivatives from Eq.~(\ref{eq:pf-dynamics2}) can be recast as

\begin{equation*}
    \partial_z \rightarrow \partial_z +\epsilon^{1/2} \partial_Z, \;\;\;
    \partial_x \rightarrow \epsilon^{1/4} \partial_X, \;\;\;
    \partial_y \rightarrow \epsilon^{1/4} \partial_Y, \;\;\;
    \partial_t \rightarrow \epsilon \partial_T
\end{equation*}

The dynamical equation for the order parameter can then be expanded in terms of these fast and slow variables. By writing its linear part as the operator $L$, we have that
\begin{eqnarray*}
  L - \partial_t
  &=& -\epsilon - \alpha(\nabla^2 + q_0^2)^2 - \partial_t \\[2mm]
  &=& -\epsilon - \alpha (
      (\partial_z+ \epsilon^{1/2} \partial_Z)
      (\partial_z+ \epsilon^{1/2} \partial_Z)
     +\epsilon^{1/2} \partial_X^2 + \epsilon^{1/2} \partial_Y^2 + q_0^2)^2
     - \epsilon \partial_T \\[2mm]
  &=& L_c + \epsilon^{1/2} L_1 + \epsilon L_2 + \epsilon^{3/2} L_3
      +\epsilon^2 L_4 \, .
\end{eqnarray*}

The phase field order parameter $\psi$ can be expanded in terms of $\epsilon$ about the zero solution as
\begin{equation*}
    \psi = \epsilon^{1/4} \psi_1 + \epsilon^{3/4} \psi_2 + \epsilon^{5/4} \psi_3
    + \ldots
\end{equation*}
By substituting these expansions back into the phase field dynamical equation, we collect the different terms according to their order in $\epsilon$. Starting with order $\epsilon^{1/4}$, we have
\begin{equation*}
    L_c \psi_1 = 0 \Rightarrow
	\psi_1(\mathbf{x},t) \;=\; \frac{1}{2} \left[ A_{11} 
    \, e^{\, i q_0 z} + {\rm c.c.} \right] \, .
\end{equation*}

For order $\epsilon^{3/4}$, the following is satisfied
\begin{equation*}
    L_c \psi_2 + L_1\psi_1 = 0 \Rightarrow
	\psi_2(\mathbf{x},t) \;=\; \frac{1}{2} \left[ A_{21} 
    \, e^{\, i q_0 z} + {\rm c.c.} \right] \, .
\end{equation*}

Finally, for order $\epsilon^{5/4}$ we find extra contributions owing to the nonlinear terms in Eq.~(\ref{eq:pf-dynamics2}),
\begin{eqnarray*}
  L_c \psi_3
  &=& - L_1 \psi_2 - L_2 \psi_1
      -\beta\psi_1^3|_{\pm i q_0}+\gamma \psi_1^5|_{\pm i q_0}
  \\[2mm]
  &=& -\bigg[-\epsilon +4\alpha q_0^2\partial_Z^2
      -i4\alpha q_0\partial_Z(\partial_X^2+\partial_Y^2)
  \\[2mm]
  && \hspace*{8mm}
     -\alpha(\partial_X^4+2\partial_X^2\partial_Y^2+\partial_Y^4)
     +\frac{3}{4}\beta |A_{11}|^2
     -\frac{5}{8}\gamma |A_{11}|^4-\partial_T\bigg]
     \left( A_{11} \, e^{\, i q_0 z} + {\rm c.c.} \right) \, .
\end{eqnarray*}

From the solvability condition (Fredholm's Alternative), this equation has a solution if
\begin{eqnarray*}
  \partial_T A_{11}
  &=& -\epsilon A_{11} +4\alpha q_0^2\partial_Z^2A_{11}
      -i4\alpha q_0\partial_Z(\partial_X^2+\partial_Y^2)A_{11}       
  \\[2mm]
  && +\frac{3}{4}\beta |A_{11}|^2A_{11}
     -\alpha(\partial_X^4+2\partial_X^2\partial_Y^2+\partial_Y^4)A_{11}
     -\frac{5}{8}\gamma |A_{11}|^4A_{11} \, .
\end{eqnarray*}

Since the fast-varying carrier is now removed from this equation,
we can reescale it back to the original variables $\{x,y,z,,t\}$. Expanding
$A$ as
\begin{equation*}
    A \;=\; \epsilon^{1/4} A_{11} + \epsilon^{3/4} A_{21} + \ldots
\end{equation*}
and going back to the original variables, we find the amplitude equation for $A$ in complex form,
\begin{equation*}
  \partial_t A \;=\; -\epsilon A + 4\alpha q_0^2 \partial_z^2 A
  - 4\,i\,\alpha q_0 \partial_z\nabla^2_{x;y} A - \alpha \nabla^4_{x,y} A
  + \frac{3}{4}\beta |A|^2A - \frac{5}{8}\gamma |A|^4A.
\end{equation*}
Although the current analysis was performed around small positive values of $\epsilon$, we observe numerically that this amplitude equation and its stationary solutions (discussed in Sec.~\ref{sec:asymp}) accurately describe the two phases and the front between them at least up to $\epsilon_c \approx 1$. 


\section{The Laplace-Beltrami operator for a curved surface}
\label{sec:lap-bel}

Let $S \subset {\rm I\!R}^3$ be a regular orientable surface, where $T_p(S)$ is the tangent plane to $S$ at $p \in S$.  Define the following sets of orthogonal frames
\begin{eqnarray*}
  \{{\bf t}_1,{\bf n},{\bf b}_1\}\; &,& \;\; {\bf t}_1 \in T_p(S) \\[3mm]
  \{{\bf t}_2,{\bf n},{\bf b}_2\}\; &,& \;\; {\bf t}_2 \neq {\bf t}_1,
                                      \;\; {\bf t}_2 \in T_p(S) \, .
\end{eqnarray*}
The differential $dN_p: T_p(S) \rightarrow T_p(S)$ of the Gauss map $N: S \rightarrow S^2$ of $S$, where ${\bf n} \in N(S)$, is a self-adjoint linear map. Therefore, for each $p \in S$ there exists an orthonormal basis $\{ {\bf t}_1,{\bf t}_2\}$ of $T_p(S)$ such that
\begin{equation*}
	dN_p({\bf t}_1) \;=\; -c_1 {\bf t}_1 \;,\;\;
    dN_p({\bf t}_2) \;=\; -c_2 {\bf t}_2 \, .
\end{equation*}
See Do Carmo~\cite{do2016differential} for a proof of this theorem.  Hence, ${\bf t}_1$ and ${\bf t}_2$ in our frames are defined as the eigenvectors at $p$, with eigenvalues (principal curvatures) $c_1$ and $c_2$. Since ${\bf t}_1$ and ${\bf t}_2$ are orthonormal, we can simply set an orthonormal frame aligned with the principal directions
\begin{equation*}
	\{{\bf t}_1(p),{\bf t}_2(p),{\bf n}(p)\} \;,\; p \in S.
\end{equation*}
Writing the surface coordinates as $s_1$ and $s_2$, we have $p = (s_1,s_2) \in S$. For a point near the surface $S$, we write the position vector as
\begin{equation*}
	{\bf r}(\lambda,s_1,s_2) = {\bf p}(s_1,s_2) + \lambda {\bf n}(s_1,s_2)
\end{equation*}
where $\lambda$ is the normal coordinate. Therefore, we obtain the following set of derivatives
\begin{eqnarray*}
  \frac{d{\bf r}}{ds_1} &=& \frac{d{\bf p}}{ds_1}
                            +\lambda\frac{d{\bf n}}{ds_1} = (1-\lambda c_1){\bf t}_1
  \\[3mm]
  \frac{d{\bf r}}{ds_2} &=& \frac{d{\bf p}}{ds_2}
                            +\lambda\frac{d{\bf n}}{ds_2} = (1-\lambda c_2){\bf t}_2
  \\[3mm]
    \frac{d{\bf r}}{d\lambda} &=& {\bf n}
\end{eqnarray*}

The covariant metric tensor (first fundamental form) can now be computed by
\begin{equation*}
	g_{ij} \;=\; <r_i,r_j> \;=\;
    \begin{bmatrix} & 1 & 0 & 0 & \\
        & 0 & (1-\lambda\,c_1)^2 & 0 & \\
        & 0 & 0 & (1-\lambda\,c_2)^2 & \end{bmatrix}.
\end{equation*}
From the orthogonality of the covariant and contravariant metric tensors, the contravariant form is
\begin{equation*}
	g_{ij}g^{ij} = \delta^i_{\, j} \; \Rightarrow \; g^{ij} \;=\;
    \begin{bmatrix} & 1 & 0 & 0 & \\
        & 0 & (1-\lambda\,c_1)^{-2} & 0 & \\
        & 0 & 0 & (1-\lambda\,c_2)^{-2} & \end{bmatrix}.
\end{equation*}

For this principal coordinate system $(\lambda,s_1,s_2)$, the infinitesimal distance with respect to a point on the surface is
\begin{equation*}
    d{\bf r} \;=\; \frac{\partial {\bf r}}{\partial \lambda}d\lambda
    + \frac{\partial {\bf r}}{\partial s_1}ds_1
    + \frac{\partial {\bf r}}{\partial s_2}ds_2
    \;=\; {\bf n}\,d\lambda + (1-\lambda c_1){\bf t}_1ds_1
    + (1-\lambda c_2){\bf t}_2ds_2.
\end{equation*}
With the metric tensor at our disposal, it is possible to obtain the Laplace-Beltrami operator for the Riemannian manifold associated with the coordinate system $(\lambda,s_1,s_2)$. The operator has the following form
\begin{equation*}
	\nabla^2 \;=\;
    \frac{1}{g^{1/2}}\partial_i\left(g^{1/2}g^{ij} \partial_j\right) \,.
\end{equation*}
where $g = \mathrm{det}({\bf g}) =(1-\lambda\,c_1)^2(1-\lambda\,c_2)^2$. We expand further as
\vspace*{2mm}
\begin{flalign*}
    \hspace*{5mm}
    &\nabla^2 \;=\; g^{ij}\partial_{ij}+\partial_i(g^{ij})\partial_j
    + \frac{1}{g^{1/2}}\partial_i(g^{1/2})g^{ij}\partial_j  \;,&
    \\[2mm]
    &g^{ij}\partial_{ij}
    \;=\; \partial_\lambda^2+(1-\lambda\,c_1)^{-2}
    \partial_{s_1}^2+(1-\lambda\,c_2)^{-2}
    \partial_{s_2}^2 \;,&
    \\[2mm]
    &\partial_i(g^{ij})\partial_j
    \;=\; \frac{2\lambda\partial_{s_1}c_1}{(1-\lambda\,c_1)^3}
    \partial_{s_1}+\frac{2\lambda\partial_{s_2}c_2}{(1-\lambda\,c_2)^3}
    \partial_{s_2} \;,&
    \\[2mm]
    & \frac{1}{g^{1/2}}\partial_i(g^{1/2})g^{ij}\partial_j
    \;=\;
    \frac{1}{g^{1/2}}\Big\{[-(c_1+c_2)+2\lambda\,c_1c_2]\partial_\lambda
    +[-\lambda\partial_{s1}(c_1+c_2)+\lambda^2\partial_{s1}(c_1c_2)]
    (1-\lambda c_1)^{-2}\,\partial_{s_1} &
    \\[1mm]
    & \hspace{43mm} + [-\lambda\partial_{s2}(c_1+c_2)+\lambda^2\partial_{s2}(c_1c_2)]
    (1-\lambda c_2)^{-2}\,\partial_{s_2} \Big\} \;.&
\end{flalign*}

For a weakly distorted interface, derivatives in the normal and the tangential direction scale differently in terms of curvatures: $\partial_\lambda \sim 1$, $\partial_{s_1} \sim c_1$ and $\partial_{s_2} \sim c_2$. Hence, by neglecting the higher order curvature contributions for tangential derivatives, the Laplace-Beltrami operator can be reduced to
\begin{eqnarray*}
  \nabla^2 &\approx& \partial_\lambda^2 + \partial_{s_1}^2+\partial_{s_2}^2+
               \frac{-(c_1+c_2)+2\lambda\,c_1c_2}
               {1-\lambda(c_1+c_2)+\lambda^2c_1c_2}\partial_\lambda
  \\[2mm]
           &=& \partial_\lambda^2 + \partial_{s_1}^2 + \partial_{s_2}^2 +
           \partial_\lambda(\mathrm{ln}
               (1-\lambda (c_1+c_2)+\lambda^2c_1c_2))\partial_\lambda .
\end{eqnarray*}
By expanding $\mathrm{ln}(1+x) = x - (1/2)x^2+(1/3)x^3+ ... $ with $x = (-2\lambda H + \lambda^2G)$, where $H = \frac{1}{2}(c_1+c_2)$ is the mean curvature and $G = c_1c_2$ the Gaussian curvature, the previous equation becomes
\begin{eqnarray}
  \nonumber
  \nabla^2
  &=& \partial_\lambda^2 + \partial_{s_1}^2+\partial_{s_2}^2
      +\partial_\lambda\bigg[-2\lambda H+\lambda^2G
  \\[2mm]
  &&\hspace{33mm}-\frac{1}{2}(4\lambda^2H^2-4\lambda^3GH+\lambda^4G^2)
     +\frac{1}{3}(-8\lambda^3H^3+...)\bigg]\partial_\lambda + h.o.t.
     \label{eq:lbexp2}
\end{eqnarray}

By rearranging the terms, Eq.~(\ref{eq:lbexp2}) may be cast with respect to its leading order terms as
\begin{equation*}
	\nabla^2 \;\approx\; \partial_\lambda^2 + \nabla^2_s +
    (-2H-(4H^2-2G)\lambda+2H(G-B)\lambda^2)\partial_\lambda
\end{equation*}
where $\nabla^2_s = \partial_{s_1}^2+\partial_{s_2}^2$. Note that $B = 4H^2-2G = (c_1^2+c_2^2)$ is known as the bending curvature, and that $2H(3G-4H^2) = -(c_1^3+c_2^3)$. We don't substitute $B$ for second order curvature term to leave the Gaussian curvature explicit in it.

The biharmonic $\nabla^4$ can similarly be expanded in curved coordinates from the Laplace-Beltrami operator in Eq.~(\ref{eq:lbexp2}). This operator is needed to derive the Gibbs-Thomson equation for the case of layers perpendicular to the interface. We collect all terms up to third order in curvatures.  We find the term $(\partial_\lambda^2+\partial_{s_1}^2+\partial_{s_2}^2)^2$ as well as additional terms associated with the first, second and third derivatives with respect to $\lambda$. As we are unable to say anything about the possible order and role of derivatives in $\lambda$, we keep all of these terms; however we keep only the lowest order term in curvature associated with each of them.  This yields,
\begin{eqnarray*}
  \nabla^4
  &\;\approx\;& (\partial_\lambda^2+\partial_{s1}^2+\partial_{s2}^2)^2
            -(2\nabla_s^2H+4H(2H^2-2G))\partial_\lambda 
  \\[2mm] &&-4(H^2-G)\partial_\lambda^2-4H (\partial_\lambda^3
             + \partial_{s_1}^2\partial_\lambda+\partial_{s_2}^2\partial_\lambda)
             -4 (\partial_{s_1} H\partial_{s_1}\partial_\lambda
             + \partial_{s_2} H\partial_{s_2}\partial_\lambda) \; .
\end{eqnarray*}


\section{Generalized Gibbs-Thomson}
\label{sec:gibbs}

In this section, we derive a generalized Gibbs-Thomson relation for the case where smectic layers are parallel to the interface.  The case where layers are perpendicular to the interface is analogous, as described in Sec.~\ref{sec:asymp}. The amplitude equation is described by Eq.~(\ref{eq:amp-gen}), and has an analytical stationary solution given by Eq.~(\ref{eq:amp-sol}) in coexistence. Our procedure for deriving a a generalized Gibbs-Thomson relation is based on the analysis by Langer for the Cahn-Hilliard model~\cite{langer1992introduction} .

The chemical potential is derived from the variational derivative of Eq.~(\ref{eq:amp-energy}) with respect to the amplitude $A$, and with $\epsilon =\epsilon_c$. Consider flat SmA planes with normal aligned to the $z$ direction, and take the front solution to be $A = A_p(z)$, as in Eq.~(\ref{eq:amp-sol}). From the discussion in Sec.~\ref{sec:asymp}, the phase of the amplitude is a constant, and the amplitude reduces to real values. Then, the chemical potential associated with a flat interface is
\begin{equation*}
    -\mu_f \;=\;
    -\epsilon A + 4\alpha q_0^2 \partial_z^2 A
    + \frac{3}{4} \beta A^3 -\frac{5}{8} \gamma A^5 .
\end{equation*}
For a curved interface situated at $\lambda_0 = 0$, the chemical potential is derived from the amplitude equation describing the evolution of a weakly curved front, in the $\{\lambda,s_1, s_2\}$ coordinate system, as detailed in Sec.~\ref{sec:asymp}. As the interface in the normal direction conserves the shape of the solution $A_p$  when the SmA layers are curved (see Fig.~\ref{fig:amp}), we consider the front to be described by $A = A_p (\lambda)$. Hence, the amplitude is aligned with the normal direction $\lambda$ to the interface. The chemical potential for the curved interface is
\begin{equation*}
	-\mu_c \;=\; -\epsilon A + 4\alpha q_0^2 \bigg[\,\partial_\lambda-2H
    -(4H^2-2G)\,\lambda
    + 2H(3G-4H^2)\,\lambda^2\bigg]\,\partial_\lambda A
    +\frac{3}{4} \beta A^3 -\frac{5}{8} \gamma A^5\;.
\end{equation*}

By multiplying both sides by the derivative of the amplitude $A$ with respect to $\lambda$ and integrating the result from a point before the transition zone (say, the smectic region) to another one after the transition zone (say, the isotropic region), we obtain
\begin{eqnarray*}
  -\int^{\infty}_{-\infty} d\lambda \,\mu_c \,\partial_\lambda A &=&
  \int^{\infty}_{-\infty} d\lambda \,\bigg\{-\epsilon A +\frac{3}{4} \beta A^3
  -\frac{5}{8} \gamma A^5
  \\[2mm]
  && \hspace*{15mm} + 4\alpha q_0^2 \Big[\,\partial_\lambda -2H
  -(4H^2-2G)\,\lambda +2H(3G-4H^2)\,\lambda^2\Big]\,\partial_\lambda A
  \bigg\} \partial_\lambda A \; .
\end{eqnarray*}
Hence, the difference between the chemical potentials of a curved and flat interface is given by
\begin{eqnarray*}
  -\int^{\infty}_{-\infty} d\lambda \,
  \partial_\lambda(\mu_c A - \mu_f A) \;=\;  
  4\alpha q_0^2 &\bigg\{&
     -2H \int^{\infty}_{-\infty}
     d\lambda\,(\partial_\lambda A)^2-(4H^2-2G)\int^{\infty}_{-\infty}
  d\lambda \,(\partial_\lambda A)^2\,\lambda
  \\[2mm]
  && +2H(3G-4H^2)\int^{\infty}_{-\infty}
  d\lambda\,(\partial_\lambda A)^2\,\lambda^2 \bigg\} \; . 
\end{eqnarray*}

The integrals on the right hand side have been defined in Sec.~\ref{sec:asymp}, Eq.~(\ref{eq:stress}), see also~\cite{murphy1966thermodynamics}.  They are the interfacial tension $\sigma_h$, the bending stress $\sigma_b$ and the torsion stress $\sigma_t$, respectively. We now write the generalized Gibbs-Thomson equation as,
\begin{equation*}
	\delta \mu \Delta A \;=\; 2H \sigma_h + (4H^2-2G)\sigma_b - 2H(3G-4H^2)\sigma_t \; .
\end{equation*}
In a similar fashion, we can derive the interface velocity equation. For this, we assume that the kinetic equation of the envelope Eq.~(\ref{eq:amp-gen}) describes a motion predominantly aligned with the normal direction {\bf n}. Recall that the interface in the normal direction conserves the shape of the solution $A_p$ for curved SmA layers (with a constant phase $\phi$), so, by the chain rule,
\begin{eqnarray*}
  \partial_\lambda A\,\partial_t {\bf r}\cdot{\bf n}
  &=& \epsilon A
      + \frac{3}{4} \beta A^3 -\frac{5}{8} \gamma A^5
      + 4\alpha q_0^2 \bigg[\,\partial_\lambda-2H
     -(4H^2-2G)\,\lambda
     +2H(3G-4H^2)\,\lambda^2\bigg]\,\partial_\lambda A \; .
\end{eqnarray*}
Since $A \approx A_p$, the right hand side of the previous equation reduces to
\begin{equation}
  \partial_\lambda A\,\partial_t {\bf r}\cdot{\bf n} \;=\; 
  4\alpha q_0^2 \bigg\{\,-2H-(4H^2-2G)
  \,\lambda+2H(3G-4H^2)\,\lambda^2\bigg\}\,\partial_\lambda A \; .
  \label{eq:vel1}
\end{equation}

Since the interface velocity $V_n$ is taken as positive when the SmA surface moves in the direction of the isotropic phase (and negative otherwise), $V_n = \partial_t {\bf r} \cdot {\bf n}$ . Then,  multiplying both sides of Eq.~(\ref{eq:vel1}) by $\partial_\lambda A$ and integrating, we obtain 
\begin{eqnarray*}
  \int^{\infty}_{-\infty}
  d\lambda\,(\partial_\lambda A)^2 \, V_n \;=\; 4\alpha q_0^2
  &\bigg\{& -2H \int^{\infty}_{-\infty} d\lambda\,(\partial_\lambda A)^2
            -(4H^2-2G) \int^{\infty}_{-\infty}
            d\lambda\,(\partial_\lambda A)^2\,\lambda
  \\[2mm]
  &&
     +2H(3G-4H^2)\int^{\infty}_{-\infty}
     d\lambda\,(\partial_\lambda A)^2\,\lambda^2 \bigg\} \; .
\end{eqnarray*}

Recalling the definitions for $\sigma_h$, $\sigma_b$ and $\sigma_t$, the interfacial velocity is
\begin{equation*}
	V_n \;=\; 4\alpha q_0^2 \bigg\{
            -2H - (4H^2-2G)\frac{\sigma_b}{\sigma_h}
           + 2H(3G-4H^2)\frac{\sigma_t}{\sigma_h} \bigg\} \; .
\end{equation*}


\section{Computational methodology}
\label{sec:comp}

We employ a hybrid spectral-finite difference scheme in space owing to the fourth-order spatial derivatives in Eq.~(\ref{eq:pf-dynamics}). All gradient terms are computed in Fourier space. Unstable or nonlinearly active modes in this model are contained in a finite band around $q_{0}$, which is an input parameter  for the model. Therefore it is possible to use controlled Fourier filtering to ensure stability of the high $q$ Fourier modes in the decomposition, and thus avoid subharmonic instability arising from modes that are strongly damped in the physical model.

All nonlinear terms are computed in real space. By using real space operations we avoid having to compute Fourier mode convolutions.  We employ a second order accurate scheme in time. Because both characteristic spatial and temporal scales derive from model parameters, it is relatively easy to maintain accuracy and stability. This is in marked contrast with the difficulties inherent in evolving macroscopic singular distributions. 

Our FFT based code solves the evolution equation for the order parameter through an in-house developed C++ code (PFSmA) which relies on the FFTW library~\cite{fftweb,frigo2005design} and standard MPI libraries for parallelization. Each core receives one to several two-dimensional slabs of real (DP) three-dimensional data sets when computing forward and inverse FFTs. The main performance bottleneck in FFT computation is communication, so the global transposition of post-processed data is a downside that compromises the parallel performance.

The PFSmA code computes the order parameter after each time step using a combination of Crank-Nicolson and Adams-Bashforth schemes in Fourier space. For such task, we define the linear operator $L_q$ and the Fourier transform $N_q$ of the nonlinear terms as
\begin{equation*}
	L_q \;=\; \omega \psi_q \;=\; -\left[\epsilon + (q^2 - q_0^2)^2)\right] \psi_q
\end{equation*}
\begin{equation*}
	N_q \;=\; \left(\beta\psi^3-\gamma\psi^5\right)_q
\end{equation*}
We then use a combination of the implicit Crank-Nicolson scheme for the linear terms with an explicit, second order Adams-Bashforth scheme for the non-linear terms in Fourier space to integrate Eq.~(\ref{eq:pf-dynamics}) and obtain $\psi$ for the new time,
\begin{equation*}
    \psi_q (t+\Delta t) \;=\;
    \frac{(1+\frac{\Delta t}{2}\omega(t))\psi_q(t)
        +\frac{\Delta t}{2}(3N_q(t-N_q(t-\Delta t))}
    {1-\frac{\Delta t}{2}\omega(t+\Delta t)}\,.
\end{equation*}

For all numerical solutions shown in this work, we use Neumann and zero normal third order derivatives as boundary conditions for the order parameter field, in order to make contact with the focal conic domains of~\cite{kim2016controlling}. In this case we use the cosine Fourier transform (DCT) for the even order derivatives of the order parameter. Our  computational domain is $\Omega = [0,L]^3$, where L is the domain length. We fix $q_0 = 1$ in all simulations, such that the grid spacing is $h = 2\pi/16q_0$, $N$ is the number of nodes  (generally $512^3$ or $1024^3$) and $L = (N^{1/3}-1)h$.


\subsection{Surface tracking and curvatures computation}

The surface is tracked by searching for points where $\psi({\bf x}) = const.$ and $|\nabla \psi({\bf x})| \neq 0$ in the transition region. Since we acquire the curvatures from this rapidly varying phase field, we need to implement an algorithm to smoothly and accurately compute them. Here, based on Megrabov's work~\cite{megrabov2014divergence}, we use the following implicit expressions
\begin{equation*}
	H \;=\; \frac{1}{2}\nabla\cdot\left(\frac{\nabla\psi}{|\nabla\psi|}\right)
\end{equation*}
\begin{equation*}
	G \;=\; -\frac{1}{2}\nabla\cdot\bigg[\nabla (\mathrm{ln} |\psi|)
    - \nabla^2\psi\frac{\nabla\psi}{|\nabla\psi|^2} \bigg] \, .
\end{equation*}

Since at each node on the mesh we are able to compute the order parameter derivatives, we rework the previous expressions to better accommodate them in the algorithm. By writing first and second derivatives of $\psi$ as $\psi_i$ and $\psi_{ij}$ respectively, where $i,j = \{x,y,z\}$, we can numerically obtain the mean and Gaussian curvatures through
\begin{eqnarray}
  H \;=\; (2|\nabla\psi|^3)^{-1}&\Big[&(\psi_y^2+\psi_z^2)\psi_{xx}
        +(\psi_x^2+\psi_z^2)\psi_{yy}
        +(\psi_x^2+\psi_y^2)\psi_{zz}
       -2(\psi_x\psi_y\psi_{xy}+\psi_x\psi_z\psi_{xz}
       +\psi_y\psi_z\psi_{yz})\Big]
  \label{eq:psih}
\end{eqnarray}
and
\begin{eqnarray}
  \nonumber
  G \;=\; |\nabla\psi|^{-4}
  &\Big\{& 
           \psi_z^2(\psi_{xx}\psi_{yy}-\psi_{xy}^2)
           +\psi_y^2(\psi_{xx}\psi_{zz}-\psi_{xz}^2)
           +\psi_x^2(\psi_{yy}\psi_{zz}-\psi_{yz}^2)
  \\[2mm]
  && +2[\psi_y\psi_{xy}(\psi_z\psi_{xz}-\psi_x\psi_{zz})
     +\psi_x\psi_{xz}(\psi_y\psi_{yz}-\psi_z\psi_{yy})
     +\psi_z\psi_{yz}(\psi_x\psi_{xy}-\psi_y\psi_{xx}) ] \Big\} \; .
  \label{eq:psig}
\end{eqnarray}


\bibliographystyle{apsrev4-1}
\bibliography{interface}

\end{document}